\documentclass[12pt, a4paper]{article}

\usepackage[english]{babel}
\usepackage[utf8]{inputenc}
\usepackage[T1]{fontenc}

\usepackage[a4paper,top=1in,bottom=1in,left=1in,right=1in,marginparwidth=1.75cm]{geometry}

\usepackage{lmodern}				  	
\usepackage[T1]{fontenc}			
\usepackage[utf8]{inputenc}			
\usepackage{indentfirst}			
\usepackage{color}					
\usepackage{graphicx}				
\usepackage{microtype} 				
\usepackage{pdflscape}				
\usepackage{natbib}					
\usepackage{amsmath}			
\usepackage{mathtools}			
\usepackage[super]{nth}			
\usepackage{setspace}
\usepackage{hyperref}
\usepackage{longtable}
\usepackage{caption}
\usepackage{subcaption}
\title{How the Availability of Higher Education Affects Incentives? Evidence from Federal University Openings in Brazil\footnote{I thank my advisor Juliano Assunção and my colleagues Maria Mittelbach, Helena Arruda, João Pedro Vieira, Julio Barros, and Tomás do Valle for comments and suggestions. All remaining errors and omissions are my sole responsibility.}}

\author{Guilherme Jardim\thanks{Departamento de Economia, PUC-Rio, Rua Marquês de São Vicente, 225/F210, Rio de Janeiro - RJ, Brazil, 22453-900. E-mail: gnjardim1@gmail.com. Declarations of interest: none.}}

\begin{document}
\maketitle
\begin{abstract}
This paper studies the impact of an university opening on incentives for human capital accumulation of prospective students in its neighborhood. The opening causes an exogenous fall on the cost to attend university, through the decrease in distance, leading to an incentive to increase effort --- shown by the positive effect on students' grades. I use an event study approach with two-way fixed effects to retrieve a causal estimate, exploiting the variation across groups of students that receive treatment at different times --- mitigating the bias created by the decision of governments on the location of new universities. Results show an average increase of $0.038$ standard deviations in test grades, for the municipality where the university was established, and are robust to a series of potential problems, including some of the usual concerns in event study models.
\end{abstract}

\vspace{0.5cm}


\textbf{Keywords}: Education, University openings, Difference-in-differences, Two-way fixed effects

\cleardoublepage
% ---
% inserir o sumario
% ---
	\tableofcontents
\cleardoublepage
% ---

% ---
% inserir lista de tabelas
% ---
	\listoftables
\cleardoublepage
% ---

% ---
% inserir lista de figuras
% ---
	\listoffigures
\cleardoublepage
% ---

% ==========================================================
% ELEMENTOS TEXTUAIS
% ==========================================================

\section*{Abbreviations}
   \begin{itemize}

\item ENEM --- National Secondary Education Examination (Exame Nacional do Ensino Médio)

\item FIES --- Higher Education Student Financing Fund (Fundo de Financiamento Estudantil)

\item IBGE --- Brazilian Institute of Geography and Statistics (Instituto Brasileiro de Geografia e Estatística)

\item IES --- Higher Education Institution (Instituição de Ensino Superior)

\item INEP --- National Institute of Educational Studies (Instituto Nacional de Estudos e Pesquisas Educacionais Anísio Teixeira)

\item MEC --- Ministry of Education (Ministério da Educação)

\item ProUni --- Federal College Voucher Program (Programa Universidade para Todos)

\item REUNI --- Federal Universities Restructuring and Expansion Plans Support Program (Programa de Apoio a Planos de Reestruturação e Expansão das Universidades Federais)

\item Sisu --- Unified Selection System (Sistema de Seleção Unificada)
\end{itemize}
\cleardoublepage

% ----------------------------------------------------------
% Introduction
% ----------------------------------------------------------
\section{Introduction}
	In 2018, there were around 200 million higher education students in the world, up from 89 million in 1998. In Latin America and the Caribbean, the number of students in higher education programs has nearly doubled in the 2000s decade \citep{world_bank_world_2018}. Between 1998 and 2018, the enrollment rates in higher education increased from 17.3\% to 38.4\% \citep{world_bank_school_2020}. In terms of institutions, the number of colleges increased from 9,103 in 2000 to 13,844 in 2013, when considering 12 countries from Latin America \citep{marta_ferreyra_at_2017}. The case of Brazil is no different, the country went from 973 higher education institutions to 2537, in the period between 1998 and 2018. In face of this scenario, it is important to understand how the availability of higher education affects local incentives. In this study, I explore the difference in timing over the placement of new universities across Brazil to investigate the immediate impact of a federal university opening on prospective students' incentives and performance. I document the effect of those openings on the academic proficiency of high school students in the neighborhood of the new university.

\begin{figure}[htbp]
	\centering
	\includegraphics[clip, trim=3.7cm 0cm 2cm 0cm]{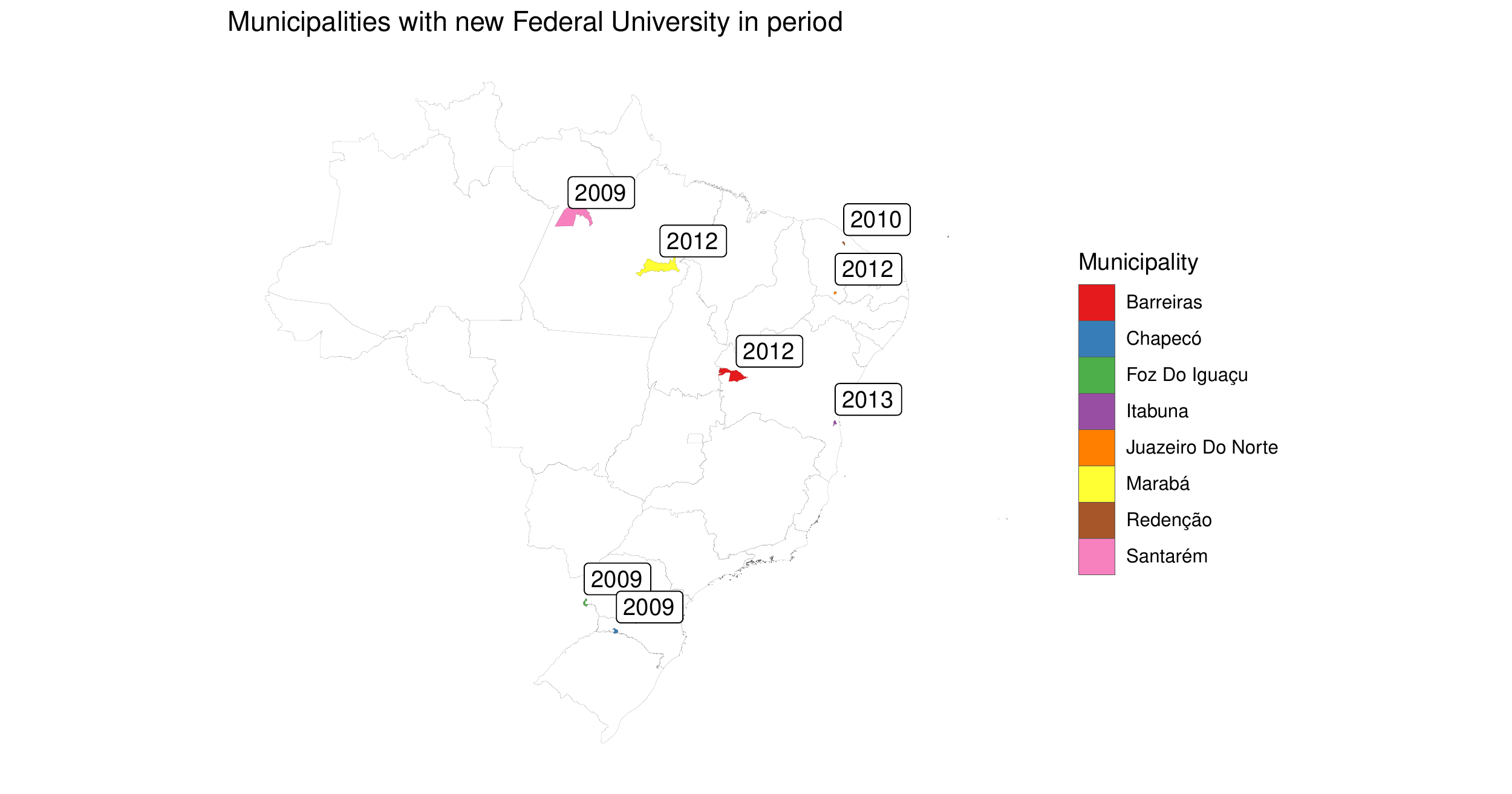}
	\caption{Federal University Openings in period}
    \label{map_munics}
\end{figure}

A better understanding on the behavior of prospective students' according to the availability of higher education can affect how public and private entities allocate universities. It also emphasizes a different perspective on how educational outcomes varies with the distribution of universities. The rationale is that when a municipality receives a new higher education institution, there is an exogenous fall on the cost to attend college, through the decrease in distance, leading to an incentive to increase effort --- which should be reflected in the grades used in the admission process. Based on a student-by-municipality-by-year panel from 2004 to 2018, I use an event study strategy to overcome the endogeneity created by the decision of the government on the placement of the institution. Specifically, the focus is in the eight federal universities founded between 2009 and 2013 in four distinct years and five different states, presented in Figure \ref{map_munics}. In a regression of test grades on exposure to a new university, controlling for socioeconomic characteristics, year and municipality fixed effects, and state-specific and controls trends, results show an average increase of $0.038$ standard deviations in test grades, for the municipality where the university was established.

The small magnitude of effects found may partially reflect the expectation that not all students are affected in the same manner. Costs faced by students vary, and the reduction caused by the decrease in distance might not have any relevant impact for the group that is not in the margin between attending or not an university. Taking into account the high estimated cost per student of U\$9.2 thousand \citep{silva_valor_2018} and the effect size, the comparison with other educational policies suggests the expansion of federal institutions combines a low cost-effectiveness with hard scalability. However, since the impact on incentives is an indirect consequence of the program, there is no clear answer regarding the overall efficacy of the policy.

The literature has focused on the effects of the opening of universities in many possible dimensions such as R\&D \citep{lehnert_employment_2020}; educational attainment \citep{currie_mothers_2003}; invention \citep{toivanen_education_2016}; migration \citep{groen_effect_2004}; participation rates and graduate outcomes \citep{frenette_universities_2009}, but there are few attempts to understand its impacts on prospective students' incentives. In addition, most previous studies that focus on the establishment of new colleges assume that their placement is random, which has been shown to overstate the effect of the opening \citep{andrews_how_2020}.

The Brazilian experience provides a good opportunity to test incentive implications caused by the establishment of a public university. Because the expansion of federal universities had the goal of reducing the geographical concentration of those in the country, many municipalities around the new university had no other federal higher education institution nearby. This allows a measurement of the effect of placing an university in a city where the transportation costs are significant and the decrease in distance is relevant. In addition, the centralization of the admission process, brought by the National Secondary Education Examination (ENEM), facilitates the comparison across different states and years, while also creating a nation-wide competitive environment, where effort is an important input for entering a federal university.

I show that the main results are robust to a series of potential problems, including some of the usual concerns in event study models. First, I show that there are no evidences of pre-existent differential trends prior to the opening of the university in the municipality. The inclusion of municipality and year fixed effects in the model also addresses concerns regarding the unbalanced nature of the sample. Second, I look for changes in composition of ENEM participants induced by the treatment. Finally, I test whether estimated impacts are due to chance.

% --- Similar applications
The analysis developed in this paper is related to the literature on higher education availability and effects on incentives. Similar applications exploring the idea of university openings as an exogenous shock in costs were developed by \cite{currie_mothers_2003}, using data on the presence of colleges in the woman's county in her 17th year as an instrumental variable for educational attainment, and \cite{bedard_human_2001}, showing that increasing university access, by expanding the university system may increase the high school dropout rate. \cite{card_using_1993} also explores the use of college proximity as an exogenous determinant of schooling, showing that the implied instrumental variables estimates of the return to schooling are 25-60\% higher than other previous ordinary least squares estimates. The relationship between distance and educational performance was studied by \cite{burde_effect_2012}, showing that enrollment rates of children fall by $16$ percentage points per mile and test scores fall by $0.19$ standard deviations per mile, and \cite{spiess_does_2010}, that concludes the distance to the nearest university at the time of completing secondary school significantly affects the decision to enroll in a university. \cite{long_college_2008} uses the average quality of colleges within a certain radius of the student as an instrument for the quality of the college at which the student attends, exploring the idea that, since there is a cost to the student of attending college far away from home, students are more likely to attend nearby. \cite{griffith_cant_2009} examine how proximity of selective schools affects the decision of whether to apply to a selective college.

% --- REUNI literature
Despite the wide range of potential impacts and broad coverage of the policy, there's still little evidence regarding the effects of the expansion of the federal universities in the period, especially for country-wide effects. \cite{aranha_programas_2012} presents an impact analysis of the expansion in the selection process of the Federal University of Minas Gerais (UFMG) from the perspective of racial inclusion, previous education and monthly family income, with a similar work in the Federal University of Pernambuco \citep{arruda_expansao_2011}. \cite{brune_instituicoes_2015} analyzes the influences on local development of the cities that host new campi of the Federal University of Paraná and Federal Technological University of Paraná.

The rest of this paper is organized as follows: Section 2 discusses the Brazilian education system, the admission process and the main higher education policies in the period from 2003 to 2014; Section 3 presents a theoretical framework for analyzing the relationship between the opening and students' incentives; Section 4 provides a description of the data; Section 5 explains the empirical strategy used to estimate the effects; Section 6 discusses the results of the paper; Section 7 assesses the robustness of estimated results; and Section 8 presents concluding remarks.

% ----------------------------------------------------------
% Institutional Background
% ----------------------------------------------------------
\section{Institutional Background}
	% --- Brazilian education system
The Brazilian education system consists of three main segments: fundamental (primary and secondary), high school and college. Fundamental education is mandatory for all citizens and lasts nine years, the high school takes three years, and college lasts between four and five years for most majors. Both private and public institutions coexist in those three cycles, with private institutions charging a fee, and the public system being free of charges at all levels. Regarding the public higher education system, universities are divided between the ones administered by the state government and the federal universities, administered by the central government, with each state having at least one public federal university. 

\cite{binelli_education_2008} show that, in general, private high schools tend to perform better in test scores than public institutions, with opposite results being seen in the higher education system --- public institutions have higher test scores, in average, than private ones. This pattern summarizes the selection bias created by the general trend: parents who are not financially constrained usually enroll their children in private schools to increase their chances of entering a public university later on. As a consequence, admission to public universities is highly competitive. Another relevant aspect of the higher education system is the geographical distribution --- most of the institutions are located in the South and Southeast regions of the country.

% --- Background of ENEM and ENEM as a selection criteria
An important element of universities' admission process is the National Secondary Education Examination (ENEM). ENEM was created in 1998 as a non-mandatory standardized national exam aiming to assess students' proficiency in four different areas --- languages, codes and related technologies; human sciences and related technologies; natural sciences and related technologies; and mathematics and its technologies --- and is realized annually by the National Institute of Educational Studies and Research (INEP). Its objective is to evaluate students' performance after the conclusion of secondary education cycle, that ranges from \nth{10} to \nth{12} grade. 

Although the participation is non-mandatory, the number of participants has increased significantly throughout the years, allowing the use of ENEM as an important indicator for the educational system in Brazil. From 2001 to 2008, the take-up rate in the whole country increased from 31.4\% to 61.8\% in public schools and from 25.21\% to 72\% in private schools \citep{camargo_information_2018}.

ENEM's importance as a selection criteria began in 2004 with the creation of the Federal College Voucher Program (ProUni), which uses ENEM's grades for selecting the beneficiaries of the program. In 2009, the exam became widely adopted as a selection criteria for university admissions, increasing its importance in the educational scenario. Until then, it was used, mostly, as a complement to the exams administered by universities for admissions. This major change follows the creation of the Unified Selection System (Sisu) in the same year, an online platform developed by the Ministry of Education, aiming to centralize the enrollment process for higher education institutions that adhered to ENEM as a selection criteria.

Also in the scope of this reformulation, proficiency measures started using the item response theory methodology --- which allows for better comparisons of performance over time --- and the multiple-choice exam was partitioned. Until 2008, the exam was composed of 63 interdisciplinary objective questions and an essay. After 2009, the test was subdivided in four areas of knowledge, each comprising 45 objective questions, along with an essay.

The changes made in 2009 brought a positive effect on students' incentives through the broadening of potential universities to attend. Before this change, admission exams were generally realized in the university's municipality, which resulted in a significant cost of transportation for non-residents, dissuading those students from applying and reducing the range of possibilities \citep{vilela_as_2017}. The unification of the admission exams in a national test, available in every state, reduces potential transportation costs, which is likely to cause an increase in students' efforts.

After 2014, the Higher Education Student Financing Fund (FIES) --- a program created in 1999, designed to fund expenses of low-income students in private higher education institutions --- also adopted a minimum ENEM grade as a requisite for its candidates, another illustration of the increase in the exam's relevance as a selection criteria over the years.

% --- Expansion of federal universities and related policies
The empirical analysis of this paper is largely motivated by the expansion of federal universities in Brazil, specifically from 2004 to 2012, driven by a set of public policies with the objective of increasing the access to higher education in the country. The two main policies related to this expansion in the period are Expansion I (Federal Universities Phase I Expansion Program), contemplating the period from 2003 to 2007, and REUNI (Federal Universities Restructuring and Expansion Plans Support Program), ranging from 2007 to 2012. 

\begin{figure}[htbp]
	\centering
	\includegraphics[clip, trim=0cm 0cm 0cm 0cm]{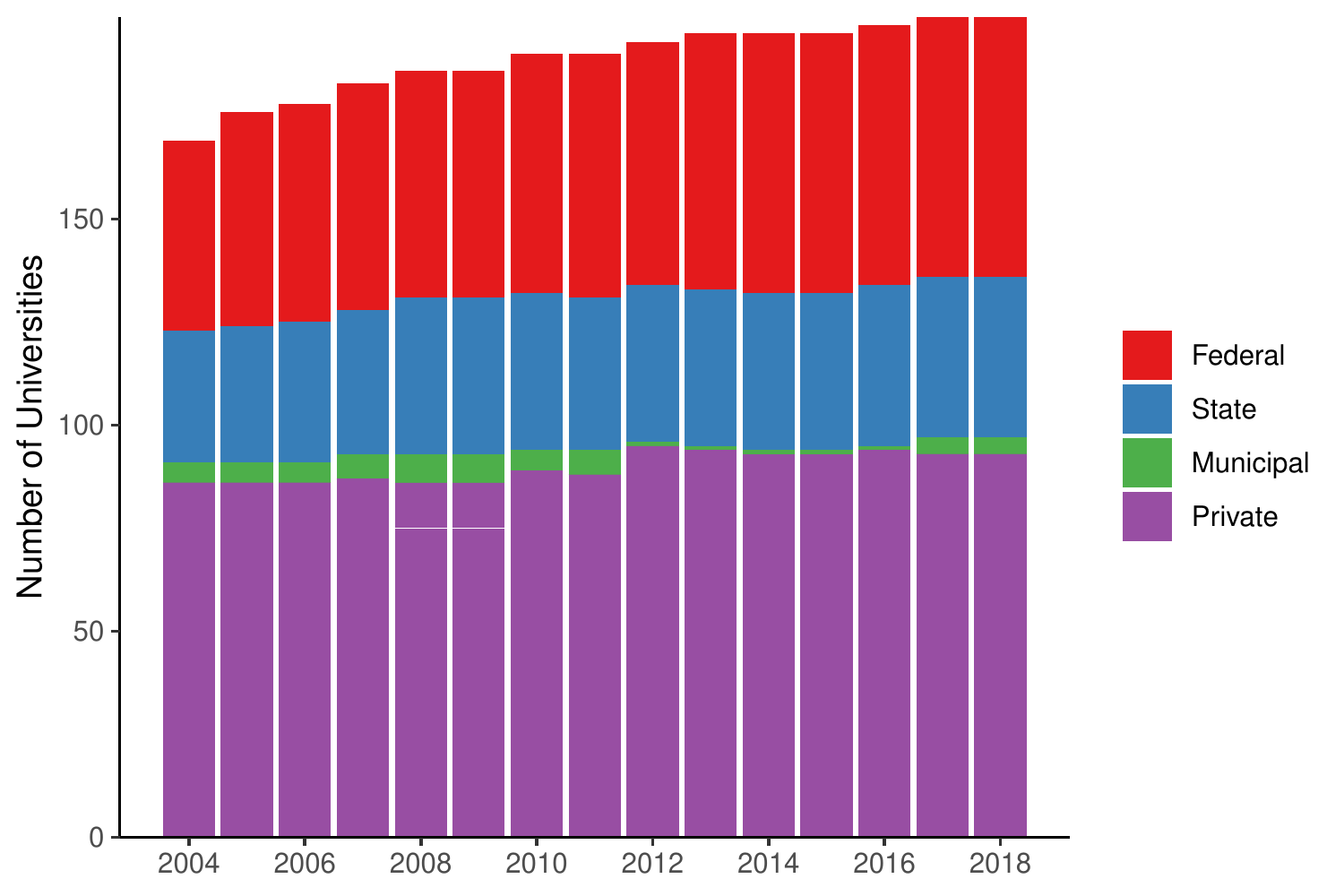}
	\caption{Evolution of the Number of Universities by Type between 2004 and 2018}
    \label{uni_types}
\end{figure}

Expansion I was created in 2003, with the goal of expanding the federal higher education to inland cities, aiming to increase the number of municipalities attended by federal universities, and was replaced by REUNI in 2007. REUNI was a program instituted by the Federal Decree 6,096 from 2007, in the scope of the Education's Development Plan, with the objective of creating conditions for the physical, academic and pedagogical expansion of the federal higher education system. Figure \ref{uni_types} shows the evolution in the number of universities in the country, detailing the composition of the sector, from 2004 to 2018. We can see that the number of State and Private institutions is roughly constant over the years, and that the Federal universities experience a relevant growth, with an increase of 28\% throughout the period between 2004 and 2012. We also observe that the federal institutions account for the majority of the public system in Brazil.

%%% Figure -----------------
\vspace*{1em}
\vspace*{1em}
%%% ------------------------

\begin{figure}[htbp]
	\centering
	\includegraphics[clip, trim=3.7cm 0cm 2cm 0cm]{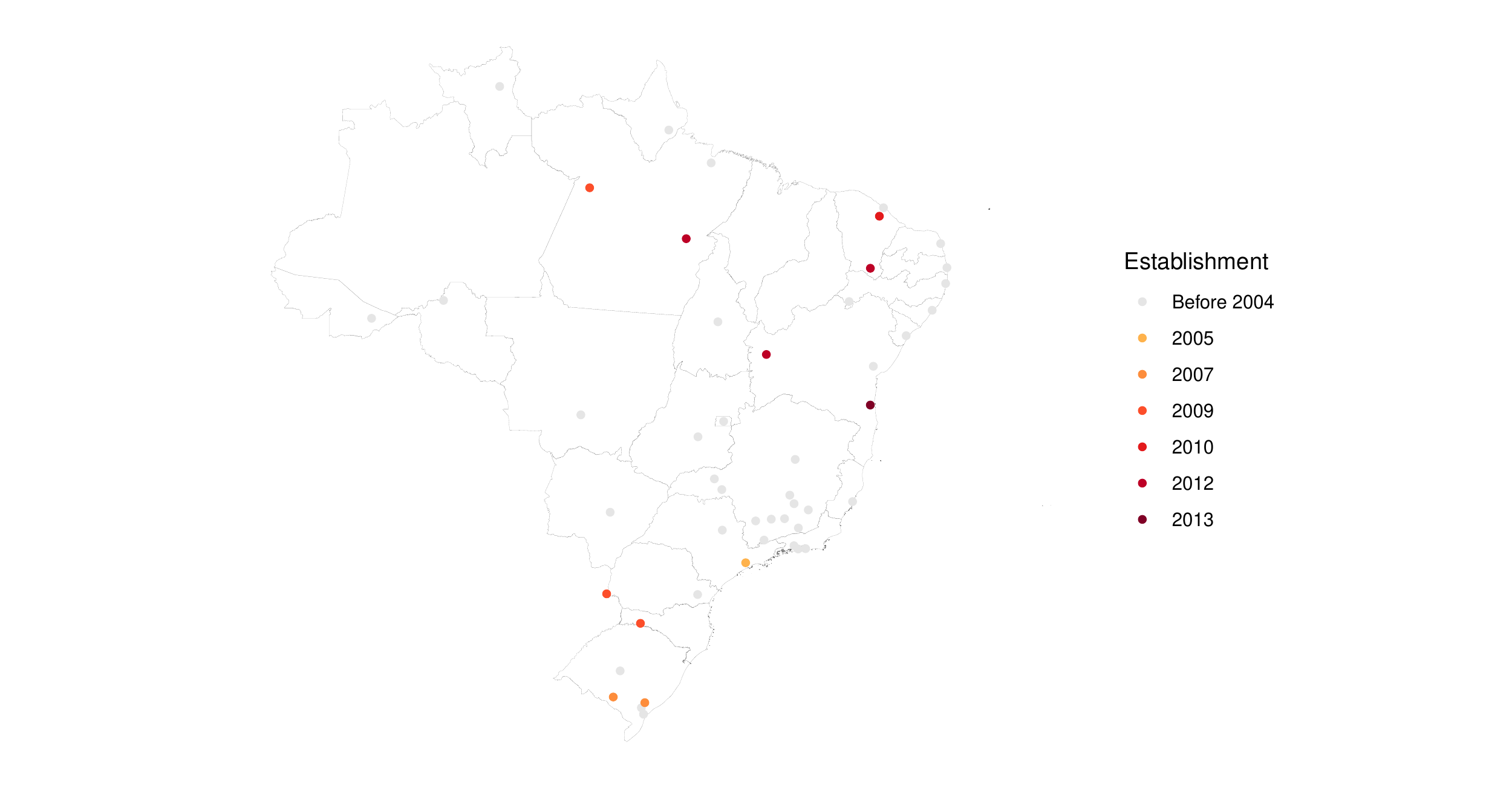}
	\caption{Municipalities with at least one Federal University by Establishment Year of the first Federal University in Municipality}
    \label{establishment_map}
\end{figure}

Figure \ref{establishment_map} shows all municipalities with at least one federal university by establishment year of the first federal university in municipality. An important aspect of the expansion carried out by the Expansion I and REUNI is the intent to increase the coverage of the higher education system across the country \citep{ministerio_da_educacao_relatorio_2012}. Despite the increase in availability of federal universities outside the South and Southeast regions of Brazil, the pattern of geographical concentration is still clear. However, through the establishment of institutions in municipalities without prior federal universities, the expansion caused a decrease on the average distance from a given municipality to a federal university, in the observed period. Therefore, the opening of a federal university changed the scenario for areas in the country that were not covered in terms of free higher education. 

%%% Figure -----------------
\vspace*{1em}
\vspace*{1em}
%%% ------------------------

\cleardoublepage

% ----------------------------------------------------------
% Theoretical Framework
% ----------------------------------------------------------
\section{Theoretical Framework}
	This paper is based on the vast theoretical and empirical literature surrounding the relationship between the costs of education and students' incentives. Starting from the seminal work of \cite{becker_investment_1962}, the predominant theoretical perspective regarding decisions about schooling is centered around the costs and returns of additional years of formal education, and my objective is to examine those decisions and costs in the context of federal university openings in Brazil. 

The human capital model proposed by Becker states that the individual should invest in education if and only if the discounted returns exceed the costs. In this setting, schooling increases productivity and, consequently, wages. Costs related to schooling are divided in two separate components: indirect, composed by foregone earnings --- the difference between what could have been earned without attending school and what is earned while in school --- and direct costs, expenses as tuition, fees, books and supplies, transportation and lodging.

%A first direct consequence of Becker's canonical model is that investments are more likely to occur when costs of schooling are lower or when returns are higher. \cite{neill_tuition_2009} explores the changes in the political party in power of different Canadian provinces to identify exogenous changes in tuition fees, measuring a large negative impact of tuition costs in enrolment. \cite{jensen_perceived_2010} estimates empirically how demand for education reacts to changes in perceived returns, noting that students given information on higher measured returns completed, on average, between $0.20$ and $0.35$ more years of school over the next four years than those who were not.

From another theoretical perspective, based on \cite{spence_job_1973}, we can understand the investment in education as a signal for potential employers in the labor market. Spence's signaling model states that schooling does not increase productivity --- unlike the premise of the human capital model --- but it acts as a signal under uncertainty. Under this model, we can interpret the costs with lodging and transportation as signaling costs and arrive in similar conclusions to the ones predicted by the human capital model. %Therefore, regardless of the chosen theoretical model, the relationship between distance-related costs and incentives is plausible, not relying on specific assumptions given by each framework.

I consider a hybrid human capital/signaling model in which attending college increases productivity and firms can identify students who graduate from university and those who do not. This information signals to firms the expected productivity of each worker. 

The probability of entering an university $\phi$ is an increasing and concave function of effort $e$. Exerting effort has a cost $c$ for the individual, an increasing and convex function of effort's level.

Firms offer a wage $\overline{\text{w}}$ if the worker graduates from university and $\underbar{w}$ otherwise, such that $\overline{\text{w}} > \underbar{w}$. If the student chooses to attend college, a cost $K$ --- associated to factors such as distance to university and tuition --- is paid. Therefore, student $i$ faces the following problem:
\begin{equation}
\max_{e_i} U = \phi(e_i) \cdot (\overline{\text{w}} - K) + [1 - \phi(e_i)] \cdot \underbar{w} - c(e_i) = 
\end{equation}
$$
= \underbar{w} + \phi(e_i) \cdot (\overline{\text{w}} - \underbar{w} - K) - c(e_i)
$$

Hence, it is clear that the optimal level of effort depends on the relationship between the wage premium $(\overline{\text{w}} - \underbar{w})$ and the costs to attend university $K$, with the first-order condition being:
\begin{equation}
\phi'(e^{*}_i) \cdot (\overline{\text{w}} - \underbar{w} - K) = c'(e^{*}_i)
\end{equation}

This can be summarized in the following proposition: the opening of an university reduces the costs associated with attending college, through the decrease in distance, which creates an incentive to increment effort in order to increase the probability of entering an university.

\section{Data}
	The empirical analysis is based on a student-by-municipality-by-year panel dataset built from multiple publicly available sources, from 2004 to 2018. Those include data provided by INEP and by the Brazilian Institute of Geography and Statistics (IBGE).

First, to determine federal university openings in each year, I use data from the Higher Education Census from 2004 to 2017. The dataset is based on a questionnaire filled by each higher education institution and data imported from the Ministry of Education (MEC) with the goal of offering detailed information regarding course, alumni, faculty and academic organization from those institutions. I can pin down when the university was established by taking the set difference $A(n) \setminus A(n-1)$, where $A(n)$ is the set of federal universities in Higher Education Census in year $n$. Thus, I build a dataset with information of municipalities that have received a federal university in the period between 2009 and 2017.

Eight federal universities were founded in this period, one in each municipality, across the states of Bahia, Ceará, Pará, Paraná and Santa Catarina, and three different regions --- South, Northeast and North. Those universities opened in four distinct years between 2009 and 2013.
 
\begin{landscape}
\begin{figure}%
    \begin{subfigure}{.5\textwidth}
	    \includegraphics[clip, trim=3.7cm 0cm 2cm 0cm, scale=0.78]{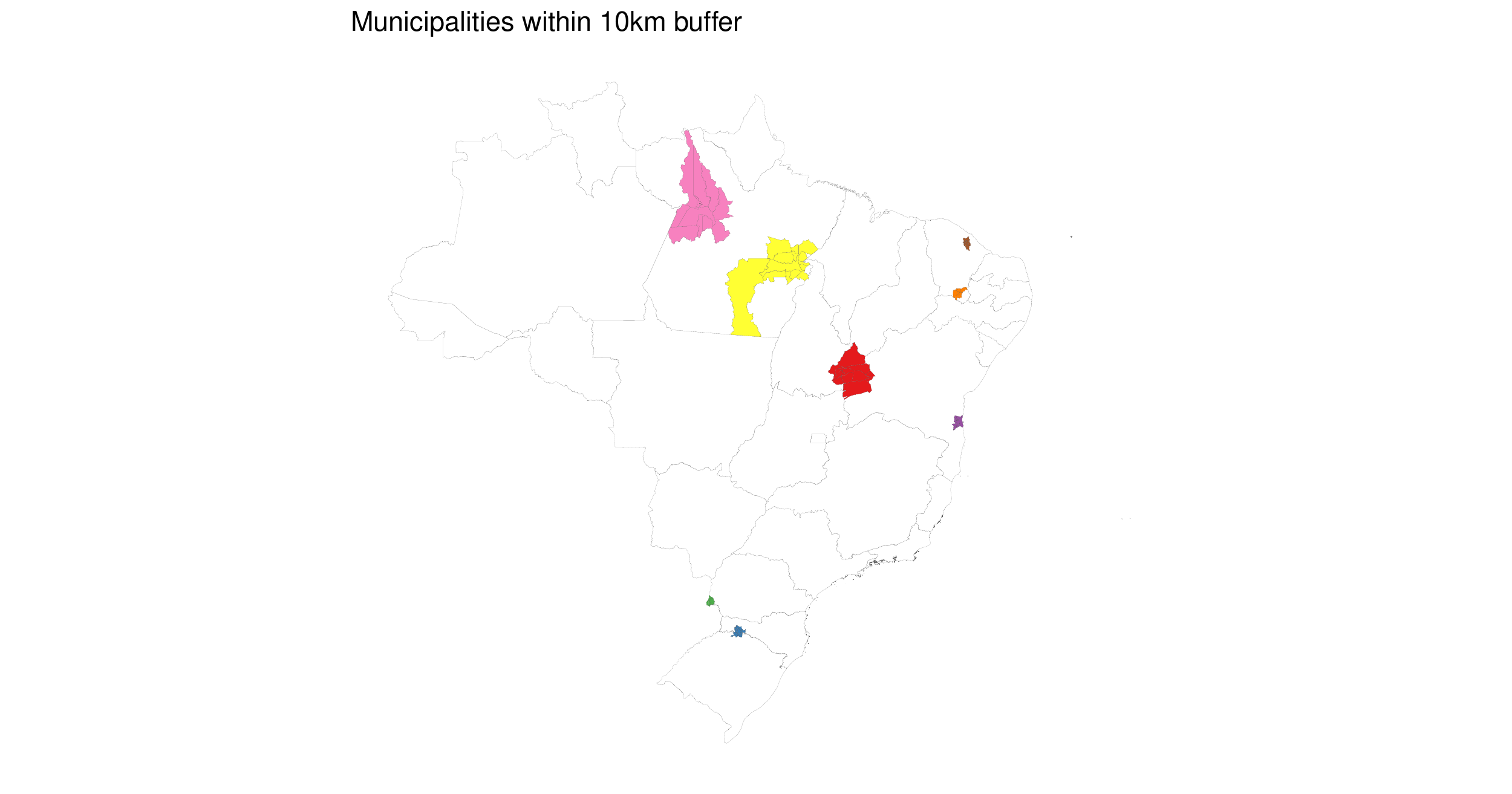}%
    \end{subfigure}
    \hspace{3cm}
    \begin{subfigure}{.5\textwidth}
		\includegraphics[clip, trim=3.7cm 0cm 2cm 0cm, scale=0.78]{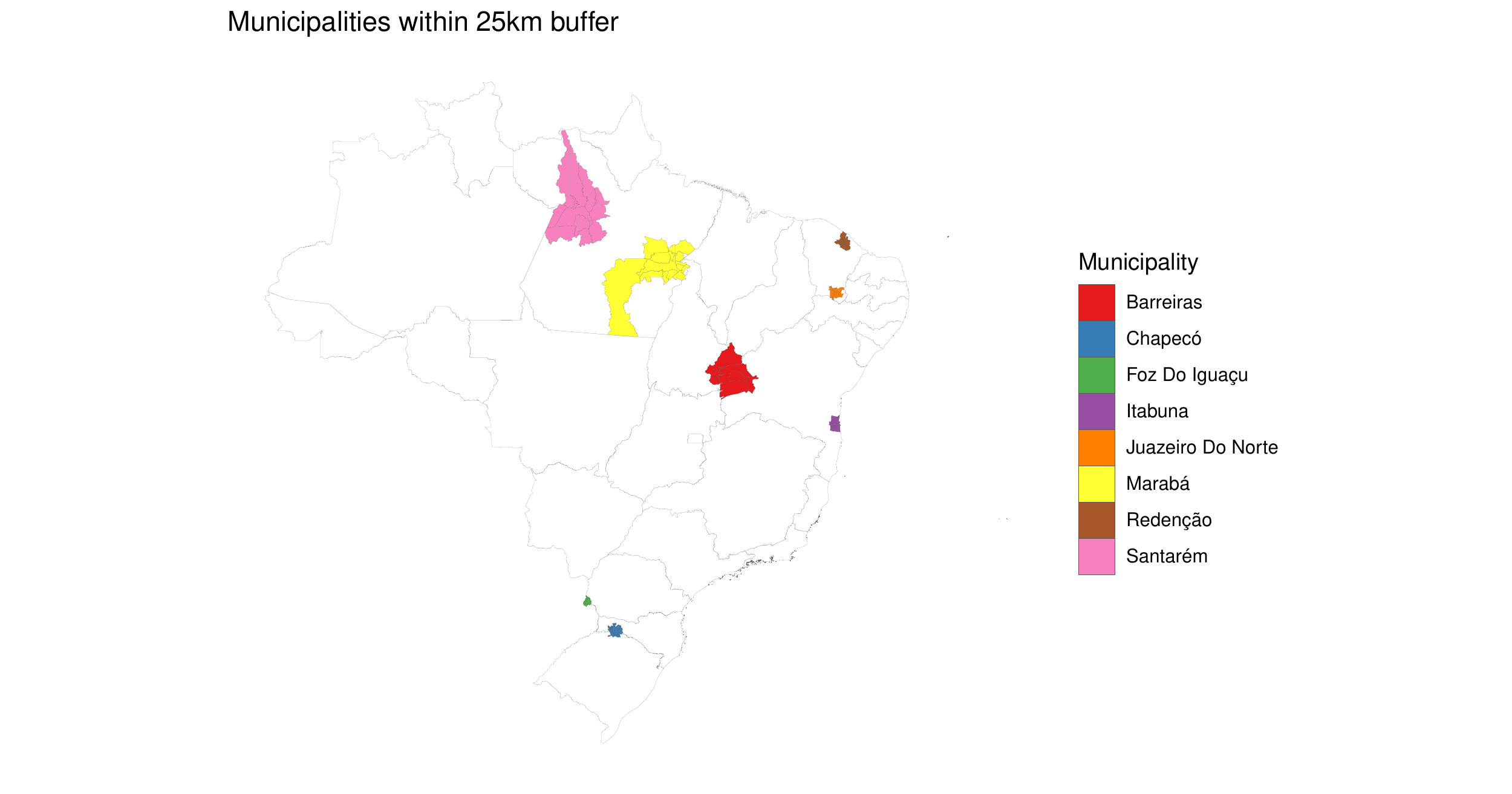}%
    \end{subfigure}
    \caption{Municipalities within each Buffer}%
    \label{fig:buffer_munis}%
\end{figure}
\end{landscape}

For the next step, I use an official spatial dataset of Brazil from IBGE and made available through the R package developed by the Institute for Applied Economic Research \citep{pereira_geobr_2019} to map out nearby municipalities using buffers of 10 and 25 kilometers from the centroid of the municipality where university was founded, those can be seen in Figure \ref{fig:buffer_munis}, by college municipality. 

I conduct the empirical analysis in all municipalities within the 25 kilometers buffer (113 in total), taking into account the possibility of heterogeneous effects across municipalities in different buffers. Since there is no overlap between buffers, I consider that no municipality is affected by more than one university simultaneously.

Next, I use data from ENEM to measure students' performance outcomes. ENEM datasets have information on students’ test scores --- in multiple-choice questions and an essay ---, as well as information on student's socioeconomic characteristics, such as age, race, gender, and family income, among others. Table \ref{tab:desc-stat} presents the distribution of socioeconomic variables considered in the study. The first column considers the eight municipalities where a federal university was established, the second comprises all municipalities within the 25km buffer, and the last considers the entire country. However, we have a lack of information related to school-level variables due to the preponderance of missing values of those in INEP's database.

For the analysis, I select students from those municipalities that took the ENEM between years of 2004 and 2018, excluding 2011 and 2012 due to incompatibility issues in microdata available. Until the present date, data available from INEP for those two years hasn't been updated to match the format of other datasets, which leads to a discrepancy in values and variables available. With that in mind, I choose to exclude those observations from the study. In addition, students who were absent in one of the two days of examination or received a zero grade in the essay were removed from the sample.

Regarding the measure of students' performance, I select the multiple-choice grade as the main outcome of interest for two reasons. First, it has homogeneous assessment criteria and doesn't involve a degree of subjectivity which may be present in essay grading. Second, it covers a much broader spectrum of knowledge and allows for a wider sampling of the content. Those factors, when combined with the discourage of guessing --- enabled by the item response theory methodology --- guarantee that the multiple-choice is a more reliable measure than the essay. Therefore, the dependent variable is defined as the average grade of the different areas assessed by the exam. 

By combining ENEM microdata at student-level with the Higher Education Census, I build a database containing examination's year and university opening year. Table \ref{tab:desc-stat} also shows the distribution of students by year and nearest municipality where university was opened. The comparison between those two dates will compose the treatment variable.

\begin{landscape}
\footnotesize 
\begin{longtable}{lllllllll}
\caption{Description of Variables} 
\label{tab:desc-stat}
\footnotesize
\\ \hline 
\hline \\[-1.8ex] 

\multicolumn{1}{c}{\textbf{Variable}} &                             & \multicolumn{2}{c}{\textbf{University   Municipalities}}      & \multicolumn{2}{c}{\textbf{Buffer Municipalities}}            & \multicolumn{2}{c}{\textbf{Whole Country}}                    &  \\ \hline
Number of Participants                &                             & 415,856                       &                               & 918,026                       &                               & 35,416,135                    &                               &  \\
\multicolumn{1}{c}{\textbf{}}         &                             & \multicolumn{1}{c}{\textbf{}} & \multicolumn{1}{c}{\textbf{}} & \multicolumn{1}{c}{\textbf{}} & \multicolumn{1}{c}{\textbf{}} & \multicolumn{1}{c}{\textbf{}} & \multicolumn{1}{c}{\textbf{}} &  \\
Grade (0--100)                         & Mean                        & 48.1                          &                               & 47.8                          &                               & 49.4                          &                               &  \\
                                      & S.d.                        & (9.5)                         &                               & (9.1)                         &                               & (10.5)                        &                               &  \\
                                      &                             &                               &                               &                               &                               &                               &                               &  \\
Gender                                & Male                        & 166,127                       & (39.9\%)                      & 367,814                       & (40.1\%)                      & 14,214,109                    & (40.1\%)                      &  \\
                                      & Female                      & 249,729                       & (60.1\%)                      & 550,212                       & (59.9\%)                      & 21,202,026                    & (59.9\%)                      &  \\
                                      &                             &                               &                               &                               &                               &                               &                               &  \\
Race                                  & White                       & 113,508                       & (27.3\%)                      & 229,816                       & (25.0\%)                      & 14,808,927                    & (41.8\%)                      &  \\
                                      & Black                       & 86,575                        & (20.8\%)                      & 184,338                       & (20.1\%)                      & 6,760,955                     & (19.1\%)                      &  \\
                                      & Pardo                       & 201,243                       & (48.4\%)                      & 470,523                       & (51.3\%)                      & 12,705,590                    & (35.9\%)                      &  \\
                                      & Yellow                      & 11,060                        & (2.7\%)                       & 24,979                        & (2.7\%)                       & 918,177                       & (2.6\%)                       &  \\
                                      & Indigene                    & 3,470                         & (0.8\%)                       & 8,370                         & (0.9\%)                       & 222,486                       & (0.6\%)                       &  \\
                                      &                             &                               &                               &                               &                               &                               &                               &  \\
Age                                   & Mean                        & 22.5                          &                               & 21.9                          &                               & 21.8                          &                               &  \\
                                      & S.d.                        & (7.4)                         &                               & (7.1)                         &                               & (7.3)                         &                               &  \\
                                      &                             &                               &                               &                               &                               &                               &                               &  \\
Father's schooling                    & No schooling                & 35,119                        & (8.4\%)                       & 86,218                        & (9.4\%)                       & 2,414,896                     & (6.8\%)                       &  \\
                                      & Elementary (years 1--5)      & 126,261                       & (30.4\%)                      & 301,526                       & (32.8\%)                      & 10,312,128                    & (29.1\%)                      &  \\
                                      & Elementary (years 6--9)      & 74,651                        & (18.0\%)                      & 167,319                       & (18.2\%)                      & 6,279,337                     & (17.7\%)                      &  \\
                                      & Incomplete high school      & 41,465                        & (10.0\%)                      & 88,597                        & (9.7\%)                       & 3,246,763                     & (9.2\%)                       &  \\
                                      & High school                 & 100,220                       & (24.1\%)                      & 206,634                       & (22.5\%)                      & 8,713,128                     & (24.6\%)                      &  \\
                                      & Incomplete higher education & 2,226                         & (0.5\%)                       & 3,743                         & (0.4\%)                       & 317,724                       & (0.9\%)                       &  \\
                                      & Higher education            & 20,903                        & (5.0\%)                       & 37,439                        & (4.1\%)                       & 2,477,463                     & (7.0\%)                       &  \\
                                      & Postgraduate                & 15,011                        & (3.6\%)                       & 26,550                        & (2.9\%)                       & 1,654,696                     & (4.7\%)                       &  \\
                                      &                             &                               &                               &                               &                               &                               &                               &  \\
Mother's schooling                    & No schooling                & 23,951                        & (5.8\%)                       & 55,235                        & (6.0\%)                       & 1,770,565                     & (5.0\%)                       &  \\
                                      & Elementary (years 1--5)      & 95,202                        & (22.9\%)                      & 231,541                       & (25.2\%)                      & 8,256,314                     & (23.3\%)                      &  \\
                                      & Elementary (years 6--9)      & 69,511                        & (16.7\%)                      & 163,093                       & (17.8\%)                      & 6,098,725                     & (17.2\%)                      &  \\
                                      & Incomplete high school      & 43,196                        & (10.4\%)                      & 96,850                        & (10.5\%)                      & 3,417,050                     & (9.6\%)                       &  \\
                                      & High school                 & 126,107                       & (30.3\%)                      & 257,551                       & (28.1\%)                      & 10,001,741                    & (28.2\%)                      &  \\
                                      & Incomplete higher education & 2,767                         & (0.7\%)                       & 4,780                         & (0.5\%)                       & 326,060                       & (0.9\%)                       &  \\
                                      & Higher education            & 28,812                        & (6.9\%)                       & 56,465                        & (6.2\%)                       & 3,056,705                     & (8.6\%)                       &  \\
                                      & Postgraduate                & 26,310                        & (6.3\%)                       & 52,511                        & (5.7\%)                       & 2,488,975                     & (7.0\%)                       &  \\
                                      &                             &                               &                               &                               &                               &                               &                               &  \\
Family income                         & No income                   & 8,185                         & (2.0\%)                       & 23,689                        & (2.6\%)                       & 738,154                       & (2.1\%)                       &  \\
                                      & 1 or less minimum wage      & 116,315                       & (28.0\%)                      & 313,131                       & (34.1\%)                      & 7,694,371                     & (21.7\%)                      &  \\
                                      & 1 to 3 minimum wage         & 201,524                       & (48.5\%)                      & 420,752                       & (45.8\%)                      & 16,260,568                    & (45.9\%)                      &  \\
                                      & 3 to 6 minimum wages        & 61,592                        & (14.8\%)                      & 113,717                       & (12.4\%)                      & 6,834,068                     & (19.3\%)                      &  \\
                                      & 6 to 9 minimum wages        & 16,645                        & (4.0\%)                       & 28,500                        & (3.1\%)                       & 2,078,794                     & (5.9\%)                       &  \\
                                      & 9 to 12 minimum wages       & 4,186                         & (1.0\%)                       & 6,747                         & (0.7\%)                       & 501,564                       & (1.4\%)                       &  \\
                                      & 12 to 15 minimum wages      & 1,861                         & (0.4\%)                       & 2,906                         & (0.3\%)                       & 237,141                       & (0.7\%)                       &  \\
                                      & More than 15 minimum wages  & 5,548                         & (1.3\%)                       & 8,584                         & (0.9\%)                       & 1,071,475                     & (3.0\%)                       &  \\
                                      &                             &                               &                               &                               &                               &                               &                               &  \\
Marital status                        & Single                      & 359,035                       & (86.3\%)                      & 806,673                       & (87.9\%)                      & 31,244,154                    & (88.2\%)                      &  \\
                                      & Married                     & 51,654                        & (12.4\%)                      & 101,424                       & (11.0\%)                      & 3,677,841                     & (10.4\%)                      &  \\
                                      & Divorced                    & 4,492                         & (1.1\%)                       & 8,743                         & (1.0\%)                       & 442,760                       & (1.3\%)                       &  \\
                                      & Widowed                     & 675                           & (0.2\%)                       & 1,186                         & (0.1\%)                       & 51,380                        & (0.1\%)                       &  \\
                                      &                             &                               &                               &                               &                               &                               &                               &  \\
Year                                  & 2004                        & 5,551                         & (1.3\%)                       & 12,080                        & (1.3\%)                       & 692,964                       & (2.0\%)                       &  \\
                                      & 2005                        & 13,707                        & (3.3\%)                       & 25,535                        & (2.8\%)                       & 1,487,482                     & (4.2\%)                       &  \\
                                      & 2006                        & 18,183                        & (4.4\%)                       & 35,194                        & (3.8\%)                       & 1,759,111                     & (5.0\%)                       &  \\
                                      & 2007                        & 16,786                        & (4.0\%)                       & 34,633                        & (3.8\%)                       & 1,916,164                     & (5.4\%)                       &  \\
                                      & 2008                        & 17,525                        & (4.2\%)                       & 33,107                        & (3.6\%)                       & 1,919,779                     & (5.4\%)                       &  \\
                                      & 2009                        & 16,145                        & (3.9\%)                       & 31,786                        & (3.5\%)                       & 1,587,942                     & (4.5\%)                       &  \\
                                      & 2010                        & 28,686                        & (6.9\%)                       & 59,855                        & (6.5\%)                       & 2,528,526                     & (7.1\%)                       &  \\
                                      & 2013                        & 47,989                        & (11.5\%)                      & 112,387                       & (12.2\%)                      & 3,678,047                     & (10.4\%)                      &  \\
                                      & 2014                        & 54,018                        & (13.0\%)                      & 125,408                       & (13.7\%)                      & 4,208,882                     & (11.9\%)                      &  \\
                                      & 2015                        & 56,038                        & (13.5\%)                      & 125,861                       & (13.7\%)                      & 4,445,594                     & (12.6\%)                      &  \\
                                      & 2016                        & 58,646                        & (14.1\%)                      & 132,146                       & (14.4\%)                      & 4,572,828                     & (12.9\%)                      &  \\
                                      & 2017                        & 41,584                        & (10.0\%)                      & 95,536                        & (10.4\%)                      & 3,525,050                     & (10.0\%)                      &  \\
                                      & 2018                        & 40,998                        & (9.9\%)                       & 94,498                        & (10.3\%)                      & 3,093,766                     & (8.7\%)                       &  \\
                                      &                             &                               &                               &                               &                               &                               &                               &  \\
College municipality                  & Marabá                      & 57,910                        & (13.9\%)                      & 132,099                       & (14.4\%)                      & \multicolumn{1}{r}{-}         &                               &  \\
                                      & Santarém                    & 96,882                        & (23.3\%)                      & 131,543                       & (14.3\%)                      & \multicolumn{1}{r}{-}         &                               &  \\
                                      & Juazeiro do Norte           & 67,711                        & (16.3\%)                      & 142,291                       & (15.5\%)                      & \multicolumn{1}{r}{-}         &                               &  \\
                                      & Redenção                    & 8,776                         & (2.1\%)                       & 179,805                       & (19.6\%)                      & \multicolumn{1}{r}{-}         &                               &  \\
                                      & Barreiras                   & 38,295                        & (9.2\%)                       & 66,623                        & (7.3\%)                       & \multicolumn{1}{r}{-}         &                               &  \\
                                      & Itabuna                     & 66,146                        & (15.9\%)                      & 143,214                       & (15.6\%)                      & \multicolumn{1}{r}{-}         &                               &  \\
                                      & Foz do Iguaçu               & 50,301                        & (12.1\%)                      & 60,432                        & (6.6\%)                       & \multicolumn{1}{r}{-}         &                               &  \\
                                      & Chapecó                     & 29,835                        & (7.2\%)                       & 62,019                        & (6.8\%)                       & \multicolumn{1}{r}{-}         &                               &  \\
                                      &                             &                               &                               &                               &                               & \multicolumn{1}{r}{}          &                               &  \\ \hline \hline
\end{longtable}

\end{landscape}

% ----------------------------------------------------------
% Empirical Strategy
% ----------------------------------------------------------
\section{Empirical Strategy}
	Since we expect that governments do not randomly choose the location of universities, simple regressions between distance to university and test scores tend to be biased. Because the placement of federal universities is expected to follow educational outcomes and demand for higher education, the effect of the college opening is likely to be overestimated.

To overcome this issue and estimate the causal impacts of new universities on student outcomes, I use a strategy of event study with two-way fixed effects estimation and follow the steps proposed by \cite{borusyak_revisiting_2017} for identification, exploiting the variation across municipalities that receive a new federal university at different times, in a setting similar to \cite{garrouste_school_2020}. Since it uses both the time and municipality dimensions, it accounts for potential selection into the treatment and time trends. The unit fixed effects control for the possibility that treated municipality have unobserved characteristics correlated with federal university openings, which implies that openings do not need to be exogenous events. In the same manner, the year fixed effects control for general changes in grades across years, possibly caused by modifications in ENEM or in admission systems.

The advantage of this method over a two-way fixed effects differences-in-differences estimator is the possibility to capture a varying treatment effect over time. In the case of the diff-in-diff strategy, this variation biases timing comparisons, resulting in the estimate being a misleading summary of the average post-treatment effect \citep{goodman-bacon_difference--differences_2018}.
 
In order to employ the event study method, I build the treatment variable as following:
\begin{equation} \label{treatment_var} 
\text {Treatment}_{m, s, k}=\left\{\begin{array}{ll}{1} & {t - \text {Opening Year}_{m, s} = k} \\ {0} & {\text {otherwise}}\end{array}\right.
\end{equation} 
where $t$ is the year in which the outcome is observed and $\text {Opening Year}_{m, s}$ is the establishment year of the nearest federal university in municipality $m$ and state $s$. Therefore, $\text {Treatment}_{m, s, k}$ denotes the number of periods relative to the event, defining one dummy variable for each year before/after the university opening. Note that I only consider the opening of a federal university.

Table \ref{tab:treat-var} presents the distribution of the treatment variable. As expected, there is a growing number of participants because the variable is positively correlated with the year, which is consistent with the greater importance attributed to the ENEM after 2009. 

\vspace{0.5cm}
\begin{table}[!htbp] 

\caption{Treatment Variable} 
\label{tab:treat-var}
\hskip-2.2cm \begin{tabular}{@{\extracolsep{10pt}}llllll} 
\\[-1.8ex]\hline 
\hline \\[-1.8ex] 

\multicolumn{1}{c}{\textbf{Variable}} &    & \multicolumn{2}{c}{\textbf{University   Municipalities}} & \multicolumn{2}{c}{\textbf{Buffer Municipalities}} \\ \hline
Years before/after college opening    & $\tau-9$ & 1,263                      & (0.3\%)                     & 2,506                   & (0.3\%)                  \\
                                      & $\tau-8$ & 5,908                      & (1.4\%)                     & 11,579                  & (1.3\%)                  \\
                                      & $\tau-7$ & 8,088                      & (1.9\%)                     & 15,796                  & (1.7\%)                  \\
                                      & $\tau-6$ & 8,330                      & (2.0\%)                     & 18,891                  & (2.1\%)                  \\
                                      & $\tau-5$ & 11,635                     & (2.8\%)                     & 26,053                  & (2.8\%)                  \\
                                      & $\tau-4$ & 14,640                     & (3.5\%)                     & 30,112                  & (3.3\%)                  \\
                                      & $\tau-3$ & 17,642                     & (4.2\%)                     & 35,438                  & (3.9\%)                  \\
                                      & $\tau-2$ & 17,906                     & (4.3\%)                     & 33,517                  & (3.7\%)                  \\
                                      & $\tau-1$ & 8,386                      & (2.0\%)                     & 15,412                  & (1.7\%)                  \\
                                      & $\tau$  & 16,708                     & (4.0\%)                     & 39,755                  & (4.3\%)                  \\
                                      & $\tau+1$  & 42,514                     & (10.2\%)                    & 81,191                  & (8.8\%)                  \\
                                      & $\tau+2$  & 30,325                     & (7.3\%)                     & 64,520                  & (7.0\%)                  \\
                                      & $\tau+3$  & 32,611                     & (7.8\%)                     & 91,849                  & (10.0\%)                 \\
                                      & $\tau+4$  & 50,416                     & (12.1\%)                    & 120,032                 & (13.1\%)                 \\
                                      & $\tau+5$  & 46,726                     & (11.2\%)                    & 108,599                 & (11.8\%)                 \\
                                      & $\tau+6$  & 42,572                     & (10.2\%)                    & 98,991                  & (10.8\%)                 \\
                                      & $\tau+7$  & 25,316                     & (6.1\%)                     & 54,308                  & (5.9\%)                  \\
                                      & $\tau+8$  & 17,880                     & (4.3\%)                     & 44,513                  & (4.8\%)                  \\
                                      & $\tau+9$  & 16,990                     & (4.1\%)                     & 24,964                  & (2.7\%)                  \\ \hline \hline
                                      
\end{tabular}
\end{table}

\vspace{0.5cm}

If I choose to include binary variables for all periods in this setting --- known as the fully dynamic specification --- the model suffers from a fundamental under-identification problem. We cannot identify the dynamic causal effects, because the passing of absolute time cannot be distinguished from relative time $k$ when there is no control group, and in presence of municipality and year fixed effects.

Therefore, I need to impose additional restrictions to the model in order to estimate treatment effects. One of the approaches developed by the authors suggests the restriction of pre-trends, which is justified when the event is unpredictable conditional on unit characteristics.

The assumption of unpredictability means the outcome cannot be affected based on anticipation of the event, thereby, there can be no pre-trends. I consider that students may anticipate the effects of the new university opening, as the public announcement is generally made 1 to 2 years before the actual foundation. Accordingly, I take into account those effects by including two leads of the treatment, representing the horizon of anticipation. 

Hence, identification assumption is the absence of pre-trends for $k < (\tau-2)$, and I proceed to test this supposition by dropping two terms $k_{1}, k_{2} < (\tau-2)$ from the fully dynamic regression, which is the minimum number of restrictions needed in order to identify the possibility of pre-trends. I choose $k_{1} = (\tau-3)$ and $k_{2} = (\tau-9)$, selecting omitted categories far apart to reduce standard errors for individual coefficients, as suggested by the authors. All regressions report standard errors clustered by Municipality to account for clustered assignment \citep{abadie_when_2017}.

Before the analysis of pre-trends, I need to define the unit characteristics to be included in the preferred specification, for which the event is unpredictable conditional on. Thus, the regression is defined as following:
\begin{equation} \label{pretrends} \begin{aligned} 
\text {Grade}_{i,m,s,t} = \alpha & + \sum_{k = \tau-8}^{\tau-4} \beta_{k} \text { Treatment}_{m, s, k} + 
\sum_{k = \tau-2}^{\tau+9} \beta_{k} \text { Treatment}_{m, s, k} + \\
& +\sum_{k = \tau-8}^{\tau-4} \rho_{k}\cdot (\text {Treatment}_{m, s, k} \times \text {Distance}_{m,s})+ \\
& +\sum_{k = \tau-2}^{\tau+9} \rho_{k}\cdot  (\text {Treatment}_{m, s, k} \times \text {Distance}_{m,s})+ \\
& + \phi_{t} + \delta_{m} + (\theta_{s} \times t) + \\ 
& + \mu_{0} \cdot {\overline{\text{Grade}}}_{-i,m,s,t} + \mu_{1} \cdot ({\overline{\text{Grade}}}_{m,s,t} \times t) + \\ 
& + \gamma_{0} \cdot X_{i,m,s,t} + \gamma_{1}\cdot (X'_{i,m,s,t} \times \text {Distance}_{m,s}) + \epsilon_{i,m,s,t}  
\end{aligned} 
\end{equation} 

In equation \ref{pretrends}, $\text {Grade}_{i,m,s,t}$ is the outcome of interest for student $i$ in municipality $m$, state $s$ and year $t$, $\text {Treatment}_{m, s, k}$ denotes the treatment variables for each number of years before/after the nearest university opening of municipality $m$, state $s$, in relative time $k$. $\phi_{t}$ and $\delta_{m}$ refer to year and municipality-specific fixed effects, respectively, and $(\theta_{s} \times t)$ represents the state-specific trends. The inclusion of interactions between the treatment variable and distance from the new federal university $(\text {Treatment}_{m, s, k} \times \text {Distance}_{m,s})$ allows for the possibility of differential treatment effects depending on the corresponding distance.

I include the variable ${\overline{Grade}}_{-i,m,s,t}$ that represents the average grade of the municipality $m$, state $s$ in year $t$ without student $i$, to account for peer-effects, and $({\overline{\text{Grade}}}_{m,s,t} \times t)$ to allow for different trends over time between municipalities with distinct average grades.

The $X_{i,m,s,t}$ corresponds to a vector of student-level controls --- such as parents' schooling, family income, race, gender --- and $(X'_{i,m,s,t} \times \text {Distance}_{m,s})$ represents a subset of those controls interacted with the distance of municipality $m$ in state $s$ from the nearest federal university opening. $\epsilon_{i,m,s,t}$ indicates the random term errors and $\alpha$, $\gamma_{0}$ , $\gamma_{1}$, $\mu_{0}$ and $\mu_{1}$ are regression coefficients to be estimated. 

The $\beta_{k}$ coefficients in this equation are used only to evaluate pre-trends, as they do not measure the effects of treatment efficiently. Due to the fact that students could anticipate the university opening after the public announcement, we should expect statistically significant impacts only for $k \geq (\tau - 2)$. The presence of any pre-trend indicates a source of endogeneity, invalidating the identification hypothesis.

\begin{landscape}
\begin{figure}[htbp]
	\centering
	\includegraphics[clip, trim=0cm 0cm 0cm 0cm]{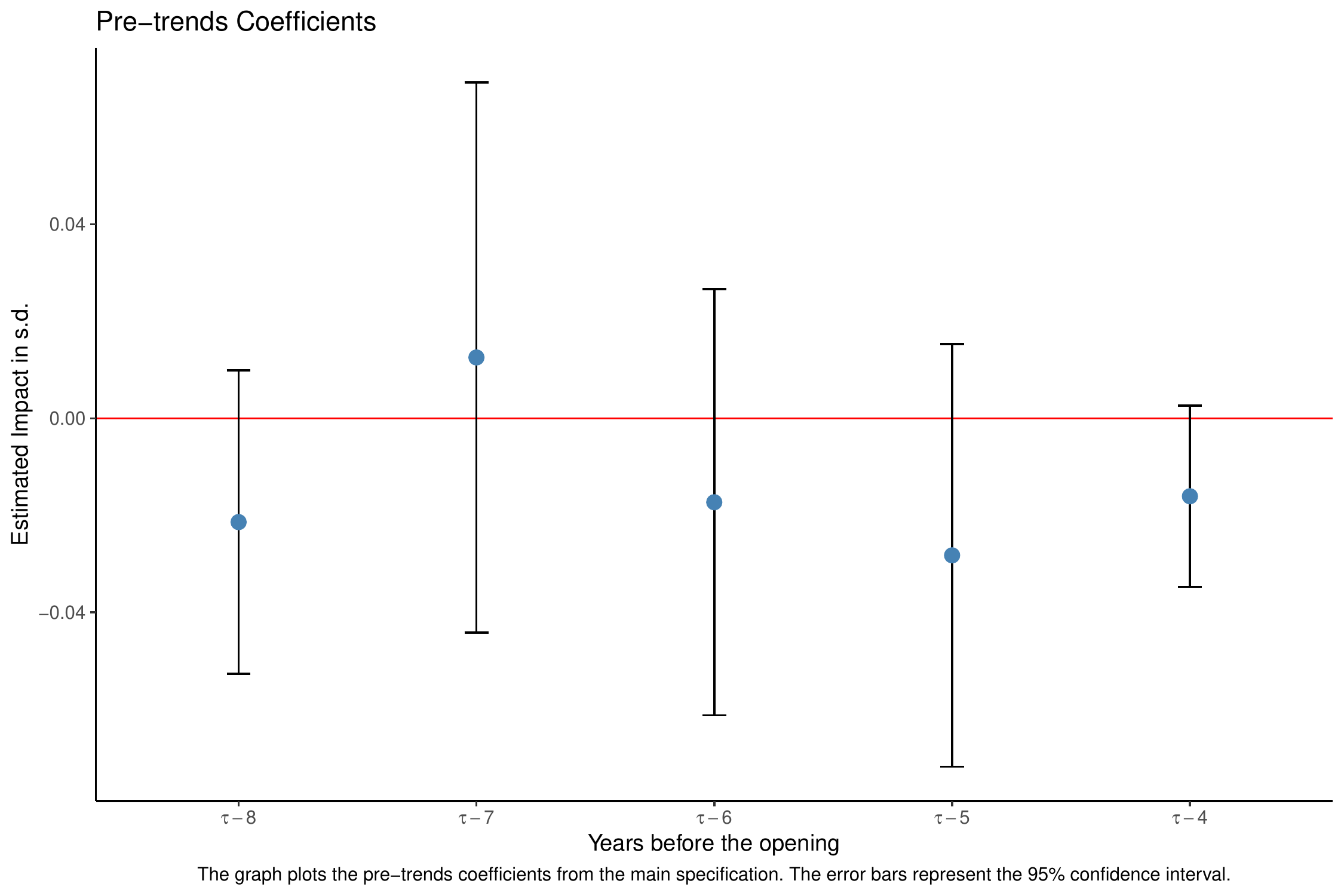}
	\caption{Pre-trends Coefficients for Regression on the Effect of University Opening on Grades}
    \label{mchoice_pretrends}
\end{figure}
\end{landscape}

Figure \ref{mchoice_pretrends} plots the coefficients of pre-trends for regression \ref{pretrends}. Results suggest the absence of pre-trends in preferred specification, with all coefficients not statistically different from zero using a 5\% significance level.

Now that we have confidence in the assumption regarding pre-trends, I set $\beta_{k} = 0, \forall k < (\tau-2)$ and estimate regressions using the semi-dynamic specification, as presented below:
\begin{equation} \label{model1} \begin{aligned} 
\text {Grade}_{i,m,s,t} = \alpha & + \sum_{k = \tau-2}^{\tau+9} \beta_{k} \text { Treatment}_{m, s, k} + \\
& +\sum_{k = \tau-2}^{\tau+9} \rho_{k} \cdot (\text {Treatment}_{m, s, k} \times \text {Distance}_{m,s})+ \\
& + \phi_{t} + \delta_{m} + (\theta_{s} \times t) + \\ 
& + \mu_{0} \cdot {\overline{\text{Grade}}}_{-i,m,s,t} + \mu_{1} \cdot({\overline{\text{Grade}}}_{m,s,t} \times t) + \\ 
& + \gamma_{0} \cdot X_{i,m,s,t} + \gamma_{1} \cdot (X'_{i,m,s,t} \times \text {Distance}_{m,s}) + \epsilon_{i,m,s,t}  
\end{aligned}  
\end{equation} 

For this regression, the binary variable indicating the time $k$ in which the nearest federal university opened for municipality $m$ is denoted by $\text {Treatment}_{m, s, k}$, with dummies ranging from $(\tau - 2)$ to $(\tau + 9)$, which allows the analysis of whether the treatment effect changes over time. Now, the $\beta_{k}$ coefficients measure the effect of treatment on student performance and can be interpreted as the cumulative impact of a new federal university at relative time $k$, compared to a baseline $(\tau - 3)$ in which the effect from university is absent, for the municipality where the university was established. The parameters $\rho_{k}$ represent the differential cumulative impact of this university at relative time $k$ for each 1 km away from the municipality where the university was founded. For instance, at a municipality contained in the 10 kilometers buffer, treatment effect at $k$ will be given by the sum $\beta_{k} + (10\cdot\rho_{k})$.

% main regression --------------------------

% ----------------------------------------------------------
% Results
% ----------------------------------------------------------
\section{Results}
	% --- Address results (see Cochrane - Writing tips for PhD students)
Table \ref{buffer_regs} presents the results for the main regressions, from equation \ref{model1}, using the standardized grade of the multiple-choice test as dependent variable and treatment dummy variables and their interactions with distance as the independent variables. Therefore, the first eight rows present the cumulative effect on the standardized test scores of a new federal university, compared to the baseline $(\tau-3)$, for a municipality where the university was established. The last eight rows show the differential cumulative impact of this university at a given period for each 1 km away from the municipality where the university was founded. The regressions include two-way fixed effects and accounts for the following controls: gender, race, age, parents' schooling, family income, marital status, and average grade of the municipality without student $i$. It also includes state-specific and control trends to allow for different trends in test grades over time between municipalities in distinct states or other characteristics. All municipalities within the 25km buffer from the new university are considered. The columns show different regression specifications, and all standard errors are clustered by municipality.

\begin{landscape}
\begin{figure}[htbp]
	\centering
	\includegraphics[clip, trim=0cm 0cm 0cm 0cm]{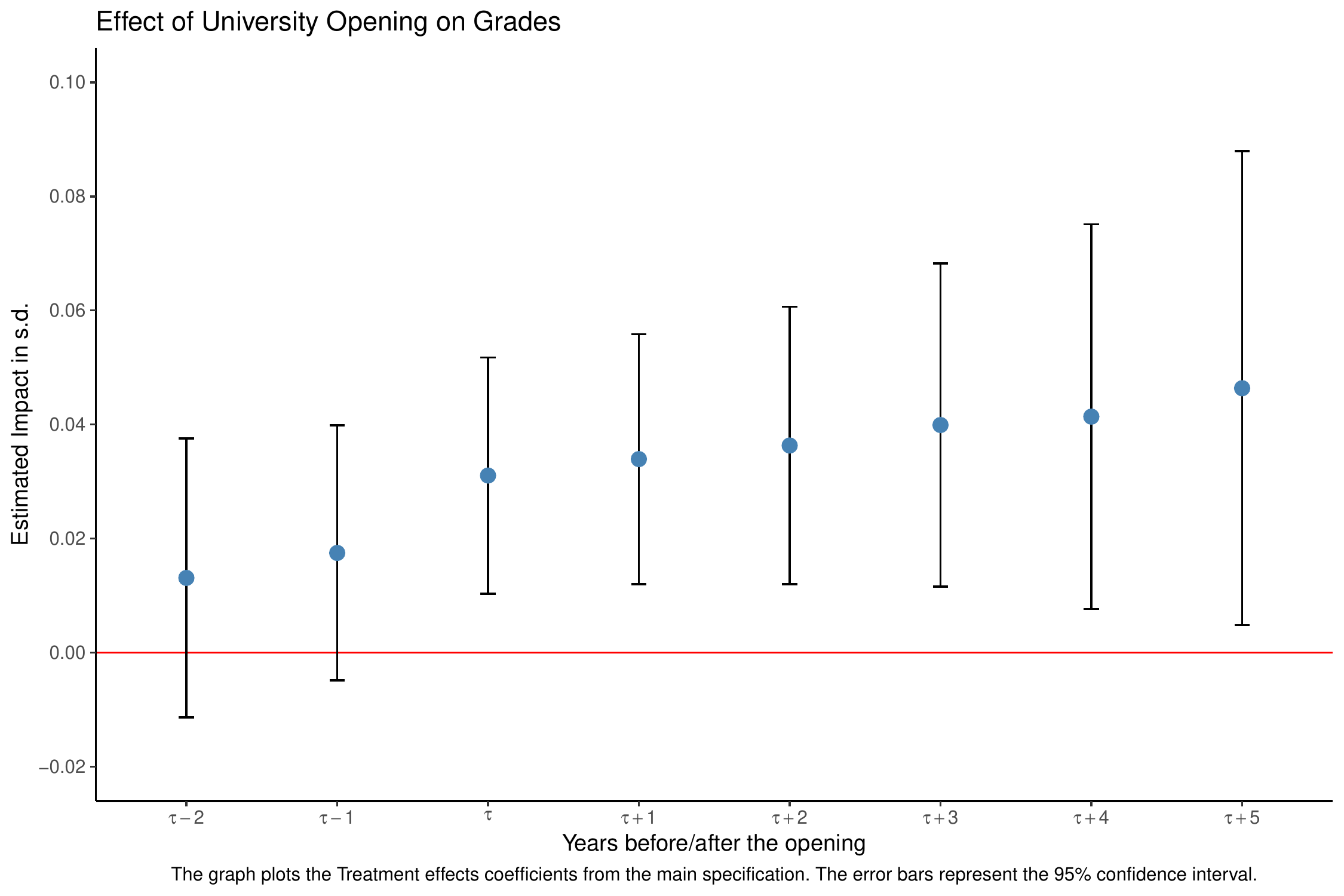}
	\caption{Treatment Coefficients for Regression on the Effect of University Opening on Grades}
    \label{mchoice_treat}
\end{figure}
\end{landscape}

The first column presents the estimated coefficients with the inclusion of fixed effects only, showing non-significant effects throughout all periods, and evidences of heterogeneous effects across distance in specific periods. The inclusion of socioeconomic and educational variables controls for individual students' characteristics, resulting in smaller standard errors and significant estimates across all years after the opening, with no signs of impact coming from the interaction between treatment and distance. Similar results are found after the addition of state-specific and control trends, with Figure \ref{mchoice_treat} plotting the treatment effects from this specification. The estimates indicate a significant effect of the federal university opening after the period $\tau$, remaining in a similar level throughout the subsequent years, with an average increase of $0.038$ standard deviation in test grades. There are evidences of heterogeneous effects in the opening year of the university with a negative coefficient indicating municipalities distant from the new federal university have a lower impact on test grades for the period $t$, with each 1km reducing the treatment effect in $0.002$ standard deviations. Comparing to the estimated effect in the period, we conclude the opening has no effect for municipalities located in the 10 km and 25 km buffers for period $\tau$.

The results suggest that individuals are constrained by the local availability of higher education, and that high school students are willing to exert more effort, in order to increase the probability of entering the university, when this constraint is alleviated and costs of schooling are lower. Therefore, the entrance of a federal university changes the scenario for areas in the country that were not covered in terms of free higher education, affecting not only students who attend the new college, but also the human capital accumulation of prospective students in its neighborhood.

% Main Regressions --------------------------
\vspace{2em}

% Table created by stargazer v.5.2.2 by Marek Hlavac, Harvard University. E-mail: hlavac at fas.harvard.edu
% Date and time: ter, set 29, 2020 - 14:12:15
\begin{table}[!htbp] \centering 
  \caption{OLS Results for the Effect of University Opening on Grades} 
  \label{buffer_regs} 
\tiny 
\begin{tabular}{@{\extracolsep{5pt}}lccc} 
\\[-1.8ex]\hline 
\hline \\[-1.8ex] 
 & \multicolumn{3}{c}{Dependent variable} \\ 
\cline{2-4} 
\\[-1.8ex] & \multicolumn{3}{c}{Standardized Grade} \\ 
\\[-1.8ex] & (1) & (2) & (3)\\ 
\hline \\[-1.8ex] 
 $\tau - 2$ & 0.038 & 0.010 & 0.013 \\ 
  & (0.052) & (0.009) & (0.012) \\ 
  & & & \\ 
 $\tau - 1$ & 0.100 & 0.009 & 0.017 \\ 
  & (0.069) & (0.015) & (0.011) \\ 
  & & & \\ 
 $\tau$ & 0.058 & 0.023$^{*}$ & 0.031$^{***}$ \\ 
  & (0.055) & (0.013) & (0.011) \\ 
  & & & \\ 
 $\tau + 1$ & 0.052 & 0.027$^{**}$ & 0.034$^{***}$ \\ 
  & (0.067) & (0.012) & (0.011) \\ 
  & & & \\ 
 $\tau + 2$ & 0.040 & 0.028$^{**}$ & 0.036$^{***}$ \\ 
  & (0.085) & (0.013) & (0.012) \\ 
  & & & \\ 
 $\tau + 3$ & 0.066 & 0.028$^{**}$ & 0.040$^{***}$ \\ 
  & (0.109) & (0.013) & (0.014) \\ 
  & & & \\ 
 $\tau + 4$ & 0.057 & 0.029$^{**}$ & 0.041$^{**}$ \\ 
  & (0.138) & (0.014) & (0.017) \\ 
  & & & \\ 
 $\tau + 5$ & 0.057 & 0.033$^{**}$ & 0.046$^{**}$ \\ 
  & (0.162) & (0.015) & (0.021) \\ 
  & & & \\ 
 $(\tau - 2)$ x buffer distance & 0.007$^{**}$ & $-$0.0001 & $-$0.0002 \\ 
  & (0.004) & (0.001) & (0.001) \\ 
  & & & \\ 
 $(\tau - 1)$ x buffer distance & 0.004 & $-$0.001 & $-$0.001$^{*}$ \\ 
  & (0.004) & (0.001) & (0.001) \\ 
  & & & \\ 
 $\tau$ x buffer distance & 0.005$^{**}$ & $-$0.001 & $-$0.002$^{**}$ \\ 
  & (0.002) & (0.001) & (0.001) \\ 
  & & & \\ 
 $(\tau + 1)$ x buffer distance & 0.004 & $-$0.001 & $-$0.001 \\ 
  & (0.003) & (0.001) & (0.001) \\ 
  & & & \\ 
 $(\tau + 2)$ x buffer distance & 0.006$^{*}$ & $-$0.0002 & $-$0.001 \\ 
  & (0.003) & (0.001) & (0.001) \\ 
  & & & \\ 
 $(\tau + 3)$ x buffer distance & 0.004 & $-$0.0002 & $-$0.001 \\ 
  & (0.003) & (0.001) & (0.001) \\ 
  & & & \\ 
 $(\tau + 4)$ x buffer distance & 0.004 & $-$0.0003 & $-$0.001 \\ 
  & (0.003) & (0.001) & (0.001) \\ 
  & & & \\ 
 $(\tau + 5)$ x buffer distance & 0.004 & 0.0003 & $-$0.001 \\ 
  & (0.003) & (0.001) & (0.001) \\ 
  & & & \\ 
\hline \\[-1.8ex] 
Year and Municipality fixed effects? & Yes & Yes & Yes \\ 
Socioeconomic and educational controls? & No & Yes & Yes \\ 
State-specific and control trends? & No & No & Yes \\ 
Number of Municipalities & 113 & 113 & 113 \\ 
Observations & 918,026 & 918,026 & 918,026 \\ 
Adjusted R$^{2}$ & 0.248 & 0.412 & 0.413 \\ 
\hline 
\hline \\[-1.8ex] 
\textit{Note:}  & \multicolumn{3}{r}{$^{*}$p$<$0.1; $^{**}$p$<$0.05; $^{***}$p$<$0.01} \\ 
\end{tabular}
\centerline{\begin{minipage}{0.95\textwidth}~\
\footnotesize{Standard errors clustered at Municipality level in parentheses. 
                          Socioeconomic and educational controls = Sex, Age, Race, Family Income, Marital status, 
                          Average grade of the municipality without student $i$, Parents' schooling} \end{minipage}} 
\end{table}

% Plots
\vspace{2em}
% -------------------------------------------

% --- Compare the magnitude of our results with other policies in education
I turn to other policies in education for a better understanding of the magnitudes of results, using findings of recent meta-analyses. Following the framework presented by \cite{627472} for causal studies evaluating effects on student achievement among upper elementary, middle and high school students, I use those effect-sizes benchmarks: less than $0.05$ is Small, $0.05$ to less than $0.20$ is Medium, and $0.20$ or greater is Large. For a more direct comparison in a closer context, \cite{camargo_information_2018} find that the public disclosure of school's average ENEM score causes an increase of test scores in 0.2 to 0.7 standard deviations for private institutions. 

Aiming for a cost-effectiveness analysis of the REUNI expansion and its effect on students' achievement, I resort to the cost estimated by \cite{silva_valor_2018} of approximately U\$9.2 thousand (R\$36.6 thousand) per student. I consider this a conservative estimate because it includes not only new universities but also expansions in existing institutions, which is expected to have lower costs when compared to university openings. We can combine this amount with the per-pupil cost benchmarks schema proposed by \cite{627472}, in which less than U\$500 is Low, U\$500 to under U\$4,000 is Moderate, and U\$4,000 or greater is High.

Those comparisons suggest the opening of federal institutions combines a low effectiveness with a high cost per student, along with a significantly hard scalability. However, a key aspect of the federal universities' expansion is that the increase in efforts, and consequently, grades is an indirect effect of the program --- its main goal is to broaden the coverage and supply of public higher education. Therefore, although it is not a sensible approach for boosting student achievement, it could be a suitable policy for other purposes.

% --- Compare with previous work on university opening
In contrast with previous works (\citealp{lehnert_employment_2020, currie_mothers_2003, toivanen_education_2016, groen_effect_2004, frenette_universities_2009}), those results represent a causal estimate of the establishment of the federal university, not relying on the assumption that their placement is random. Beyond that, the present analysis focuses on a short-term evaluation regarding the opening of a new university, considering the dimension of high school students' incentives --- measured by their performance.

% ----------------------------------------------------------
% Robustness
% ----------------------------------------------------------
\section{Robustness}
	In order to check the robustness of results to changes in model specification, I assess the possibility of anticipation regarding university openings and the prospect of estimated effects being due to chance. 

A potential threat to the model comes from the possibility of changes in the composition of participants, induced by the announcement of the university opening, just before treatment --- which could indicate an anticipation of the events by a determined group of students, meaning the assumption of unpredictability would not hold. In this case, there would be a discontinuity in allocations and it would not be possible to separate the treatment effect from the modification in treated population. If those students have unobserved characteristics correlated to higher grades, the results would be overestimated. As a test, I compare the composition of participants across treatment periods, presented in Table \ref{composition_regs}, with respect to observable characteristics. The results indicate there is no discontinuity or significant changes in the composition, which increases the confidence regarding the estimated impacts of the university openings on grades. 

% Composition Regressions -------------------
\begin{landscape}

% Table created by stargazer v.5.2.2 by Marek Hlavac, Harvard University. E-mail: hlavac at fas.harvard.edu
% Date and time: qui, set 17, 2020 - 14:28:27
\begin{table}[!htbp] \centering 
  \caption{OLS Results for the Effect of University Opening on Social Composition of Participants} 
  \label{composition_regs} 
\tiny 
\begin{tabular}{@{\extracolsep{5pt}}lcccccc} 
\\[-1.8ex]\hline 
\hline \\[-1.8ex] 
 & \multicolumn{6}{c}{Dependent variable} \\ 
\cline{2-7} 
\\[-1.8ex] & Male & White & Age & Father completed high school & Mother completed high school & Family Income > 6 minimum wages \\ 
\\[-1.8ex] & (1) & (2) & (3) & (4) & (5) & (6)\\ 
\hline \\[-1.8ex] 
 $\tau - 2$ & $-$0.003 & 0.008 & 0.002 & 0.001 & 0.0003 & 0.010 \\ 
  & (0.004) & (0.007) & (0.154) & (0.007) & (0.008) & (0.006) \\ 
  & & & & & & \\ 
 $\tau - 1$ & 0.006 & $-$0.005 & $-$0.180 & 0.009 & 0.008 & 0.008 \\ 
  & (0.005) & (0.006) & (0.287) & (0.012) & (0.012) & (0.007) \\ 
  & & & & & & \\ 
 $\tau$ & 0.006 & 0.003 & 0.342 & 0.010 & 0.003 & 0.015 \\ 
  & (0.005) & (0.009) & (0.296) & (0.014) & (0.015) & (0.010) \\ 
  & & & & & & \\ 
 $\tau + 1$ & 0.008 & 0.002 & 0.510 & $-$0.004 & $-$0.023 & 0.013 \\ 
  & (0.006) & (0.014) & (0.442) & (0.016) & (0.018) & (0.015) \\ 
  & & & & & & \\ 
 $\tau + 2$ & 0.006 & 0.007 & 0.124 & 0.001 & $-$0.024 & 0.017 \\ 
  & (0.007) & (0.015) & (0.560) & (0.019) & (0.022) & (0.020) \\ 
  & & & & & & \\ 
 $\tau + 3$ & 0.007 & 0.005 & 0.028 & 0.011 & $-$0.011 & 0.025 \\ 
  & (0.008) & (0.016) & (0.626) & (0.022) & (0.027) & (0.024) \\ 
  & & & & & & \\ 
 $\tau + 4$ & 0.005 & $-$0.002 & 0.162 & 0.009 & $-$0.021 & 0.022 \\ 
  & (0.009) & (0.018) & (0.701) & (0.027) & (0.033) & (0.029) \\ 
  & & & & & & \\ 
 $\tau + 5$ & $-$0.003 & $-$0.002 & 0.258 & 0.007 & $-$0.030 & 0.017 \\ 
  & (0.010) & (0.021) & (0.786) & (0.031) & (0.038) & (0.035) \\ 
  & & & & & & \\ 
 Intercept & 0.328$^{***}$ & 0.250$^{***}$ & 18.238$^{***}$ & 0.209$^{***}$ & 0.362$^{***}$ & 0.134$^{***}$ \\ 
  & (0.006) & (0.010) & (0.269) & (0.012) & (0.014) & (0.023) \\ 
  & & & & & & \\ 
\hline \\[-1.8ex] 
Year and Municipality fixed effects? & Yes & Yes & Yes & Yes & Yes & Yes \\ 
Number of Municipalities & 113 & 113 & 113 & 113 & 113 & 113 \\ 
Observations & 918,026 & 918,026 & 918,026 & 918,026 & 918,026 & 918,026 \\ 
Adjusted R$^{2}$ & 0.005 & 0.194 & 0.048 & 0.034 & 0.030 & 0.039 \\ 
\hline 
\hline \\[-1.8ex] 
\textit{Note:}  & \multicolumn{6}{r}{$^{*}$p$<$0.1; $^{**}$p$<$0.05; $^{***}$p$<$0.01} \\ 
\end{tabular}
\centerline{\begin{minipage}{0.95\textwidth}~\
\footnotesize{Standard errors clustered at Municipality level in parentheses.} \end{minipage}} 
\end{table} 

\end{landscape}
% --------------------------------------------

Another question related to robustness, addressed by the model, is the unbalanced nature of the sample --- not all units appear for the same number of periods before and after the initial treatment. In this setting, a problem could arise if we have evidences of a correlation between the time of treatment and the unit fixed effects. Because the sample is unbalanced, they would spend a bigger share of the sample under treated status, causing the estimated coefficients on the treatment variables to partly reflect a selection bias. As showed by \cite{borusyak_revisiting_2017}, the introduction of municipality fixed effects should address the changing composition of the sample, allowing for a unbiased estimate, even though we expect early-treated municipalities to have distinct characteristics.

\begin{landscape}
\begin{figure}[htbp]
	\centering
	\includegraphics[clip, trim=0cm 0cm 0cm 0cm]{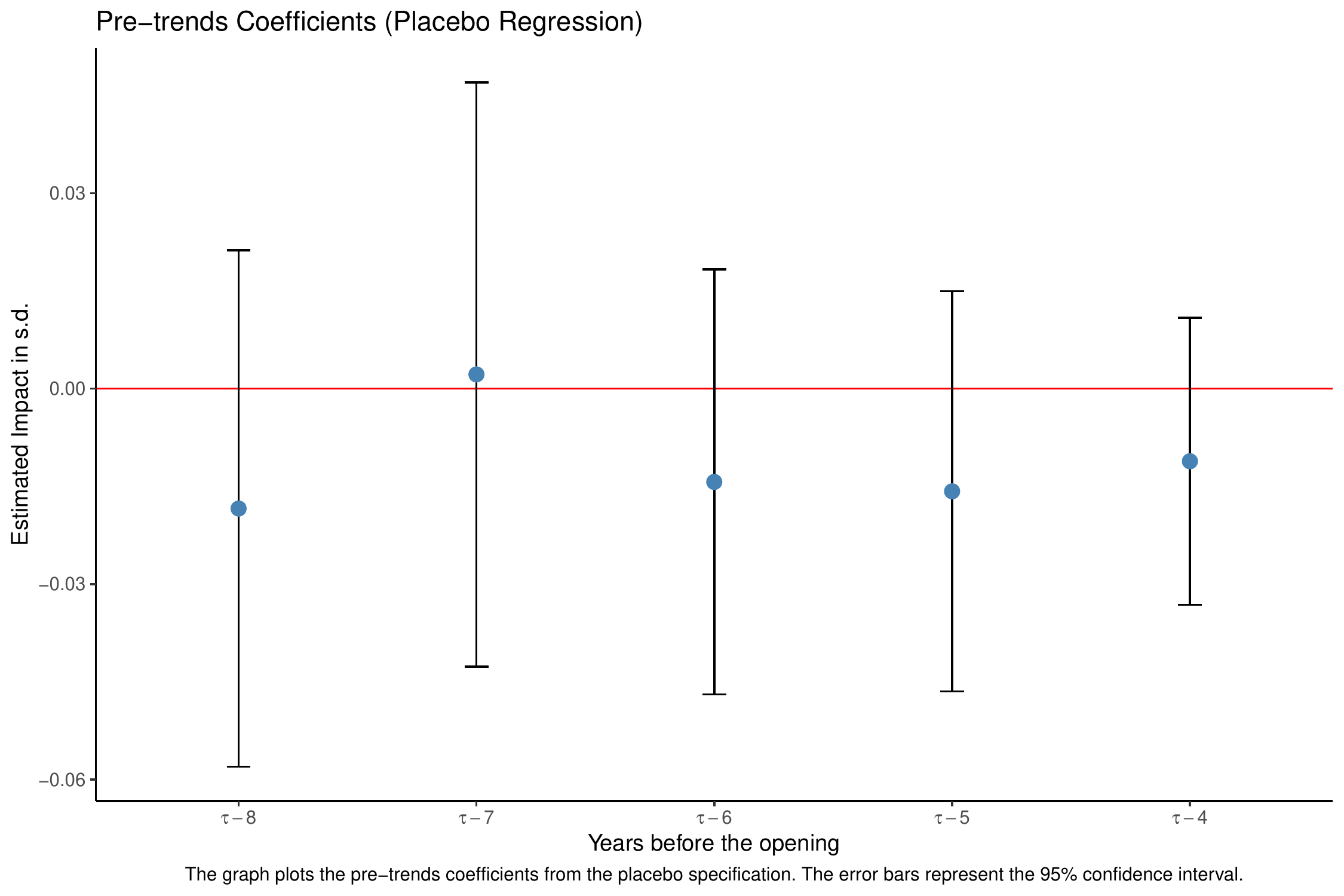}
	\caption{Pre-trends Coefficients for Regression on the Effect of University Opening on Grades -- Placebo}
    \label{robust_mchoice_pretrends}
\end{figure}
\end{landscape}

\begin{landscape}
\begin{figure}[htbp]
	\centering
	\includegraphics[clip, trim=0cm 0cm 0cm 0cm]{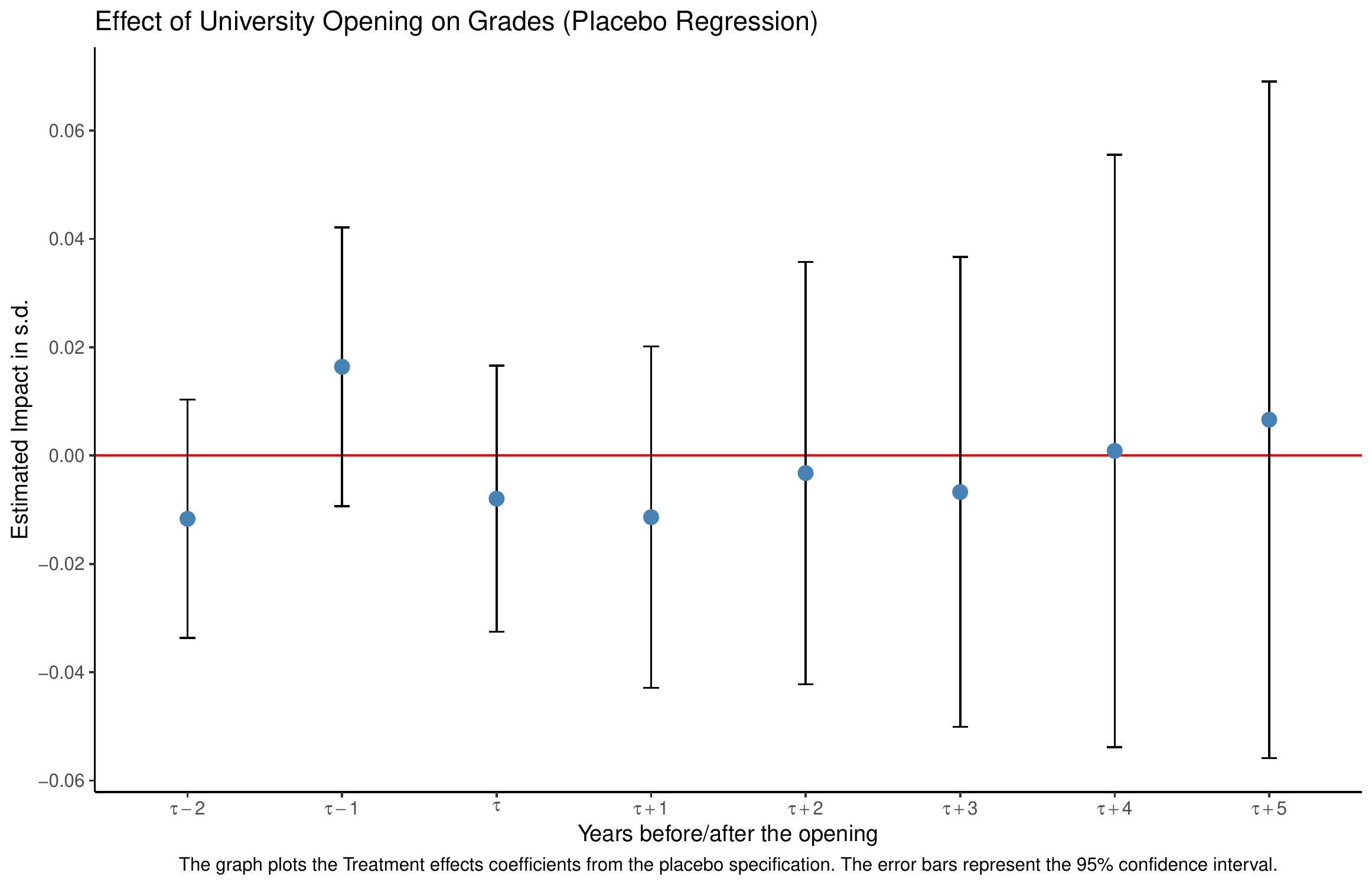}
	\caption{Treatment Coefficients for Regression on the Effect of University Opening on Grades -- Placebo}
    \label{robust_mchoice_treat}
\end{figure}
\end{landscape}

In order to test whether the estimated effects are due to chance, I run placebo regressions using municipalities between the 25km buffer and the 50km buffer. The expectation is that those municipalities have some similar characteristics to the treated ones, due to its closeness, but the distance to university suggests that the opening should have little or no effect in those cities. I use the same event study regressions from equation \ref{model1} in this subset to confirm that there are no impacts of the event, with pre-trends presented in Figure \ref{robust_mchoice_pretrends}, and results presented in Figure \ref{robust_mchoice_treat}. Since there are no significant effects on grades of the university openings in those municipalities after the inclusion of two-way fixed effects, controls and trends, I assume the effects found are genuine.

% Placebo pre-trends ------------------

% Placebo treatment coefficients -------

% ----------------------------------------------------------
% Conclusion
% ----------------------------------------------------------
\section{Conclusion}
	In this study, I explore the difference in timing over the placement of new universities across Brazil to investigate the immediate impact of the opening of a federal university on prospective students' incentives and performance. The rationale is that when a municipality receives a new higher education institution, there is an exogenous fall on the cost to attend university, through the decrease in distance, leading to an incentive to increase effort --- which should be reflected in the grades used in admission process. I use an event study approach with two-way fixed effects to retrieve a causal estimate, exploiting the variation across groups of students that receive treatment at different times --- mitigating the bias created by the decision of governments on the location of new universities. 

Results show an average increase of $0.038$ standard deviations in test grades, for the municipality where the university was established --- small but significant effects of the university openings for, at least, 5 years after the establishment. Estimates are robust to differential treatment effects over time and over distance to university, and unbalanced samples. I interpret those findings as a response of students to the reduction in distance-related costs, which is reflected in incentives to increase effort and preparation time for the admission exam. The magnitude of effects can, at least partially, be explained by the unfocused aspect of the intervention --- since we expect only the subset of students in the margin between attending or not a federal university to be affected, the average impact is small. 

Findings suggest that individuals are constrained by the local availability of higher education, and that prospective students are willing to exert more effort, in order to increase the probability of entering the university, when this constraint is alleviated and costs of schooling are lower. Therefore, the entrance of a federal university changes the scenario for areas in the country that were not covered in terms of free higher education, affecting not only students who attend the new college, but also the human capital accumulation of prospective students in its neighborhood.

% ----------------------------------------------------------
% References
% ----------------------------------------------------------

\bibliographystyle{apalike}
\bibliography{higher-education_incentives}

\newcommand{\noopsort}[1]{}
\begin{thebibliography}{}

\bibitem[Abadie et~al., 2017]{abadie_when_2017}
Abadie, A., Athey, S., Imbens, G.~W., and Wooldridge, J. (2017).
\newblock When {{Should You Adjust Standard Errors}} for {{Clustering}}?
\newblock {\em Working Paper}.

\bibitem[Andrews, 2020]{andrews_how_2020}
Andrews, M. (2020).
\newblock How {{Do Institutions}} of {{Higher Education Affect Local
  Invention}}? {{Evidence}} from the {{Establishment}} of {{U}}.{{S}}.
  {{Colleges}}.
\newblock {{SSRN Scholarly Paper}} ID 3072565, {Social Science Research
  Network}, {Rochester, NY}.

\bibitem[Aranha et~al., 2012]{aranha_programas_2012}
Aranha, A. V.~S., Pena, C.~S., and Ribeiro, S. H.~R. (2012).
\newblock {Programas de inclus\~ao na UFMG: o efeito do b\^onus e do Reuni nos
  quatro primeiros anos de vig\^encia - um estudo sobre acesso e
  perman\^encia}.
\newblock {\em Educa\c{c}\~ao em Revista}, 28(4):317--345.

\bibitem[{\noopsort{arruda}}de Arruda, 2011]{arruda_expansao_2011}
{\noopsort{arruda}}de Arruda, A. L.~B. (2011).
\newblock {Expans\~ao da educa\c{c}\~ao superior: uma an\'alise do programa de
  apoio a planos de reestrutura\c{c}\~ao e expans\~ao das Universidades
  Federais (REUNI) na Universidade Federal de Pernambuco}.
\newblock https://repositorio.ufpe.br/handle/123456789/3825.

\bibitem[Becker, 1962]{becker_investment_1962}
Becker, G.~S. (1962).
\newblock Investment in {{Human Capital}}: {{A Theoretical Analysis}}.
\newblock {\em Journal of Political Economy}, 70(5):9--49.

\bibitem[Bedard, 2001]{bedard_human_2001}
Bedard, K. (2001).
\newblock Human {{Capital}} versus {{Signaling Models}}: {{University Access}}
  and {{High School Dropouts}}.
\newblock {\em Journal of Political Economy}, 109(4):749--775.

\bibitem[Binelli et~al., 2008]{binelli_education_2008}
Binelli, C., Meghir, C., and {Menezes-Filho}, N.~A. (2008).
\newblock Education and wages in {{Brazil}}.
\newblock {\em Working Paper}.

\bibitem[Borusyak and Jaravel, 2017]{borusyak_revisiting_2017}
Borusyak, K. and Jaravel, X. (2017).
\newblock Revisiting {{Event Study Designs}}.
\newblock {{SSRN Scholarly Paper}} ID 2826228, {Social Science Research
  Network}, {Rochester, NY}.

\bibitem[Br{\"u}ne, 2015]{brune_instituicoes_2015}
Br{\"u}ne, S. (2015).
\newblock {Institui\c{c}\~oes de ensino superior e desenvolvimento: o caso do
  programa REUNI}.

\bibitem[Burde and Linden, 2012]{burde_effect_2012}
Burde, D. and Linden, L.~L. (2012).
\newblock The {{Effect}} of {{Village}}-{{Based Schools}}: {{Evidence}} from a
  {{Randomized Controlled Trial}} in {{Afghanistan}}.
\newblock Working {{Paper}} 18039, {National Bureau of Economic Research}.

\bibitem[Camargo et~al., 2018]{camargo_information_2018}
Camargo, B., Camelo, R., Firpo, S., and Ponczek, V. (2018).
\newblock Information, {{Market Incentives}}, and {{Student Performance
  Evidence}} from a {{Regression Discontinuity Design}} in {{Brazil}}.
\newblock {\em Journal of Human Resources}, 53(2):414--444.

\bibitem[Card, 1993]{card_using_1993}
Card, D. (1993).
\newblock Using {{Geographic Variation}} in {{College Proximity}} to
  {{Estimate}} the {{Return}} to {{Schooling}}.
\newblock {{SSRN Scholarly Paper}} ID 420302, {Social Science Research
  Network}, {Rochester, NY}.

\bibitem[Currie and Moretti, 2003]{currie_mothers_2003}
Currie, J. and Moretti, E. (2003).
\newblock Mother's {{Education}} and the {{Intergenerational Transmission}} of
  {{Human Capital}}: {{Evidence}} from {{College Openings}}.
\newblock {\em The Quarterly Journal of Economics}, 118(4):1495--1532.

\bibitem[Frenette, 2009]{frenette_universities_2009}
Frenette, M. (2009).
\newblock Do universities benefit local youth? {{Evidence}} from the creation
  of new universities.
\newblock {\em Economics of Education Review}, 28(3):318--328.

\bibitem[Garrouste and Zaiem, 2020]{garrouste_school_2020}
Garrouste, M. and Zaiem, M. (2020).
\newblock School supply constraints in track choices: {{A~French}} study using
  high school openings.
\newblock {\em Economics of Education Review}, 78:102041.

\bibitem[{Goodman-Bacon}, 2018]{goodman-bacon_difference--differences_2018}
{Goodman-Bacon}, A. (2018).
\newblock Difference-in-{{Differences}} with {{Variation}} in {{Treatment
  Timing}}.
\newblock {\em Working Paper}, page w25018.

\bibitem[Griffith and Rothstein, 2009]{griffith_cant_2009}
Griffith, A.~L. and Rothstein, D.~S. (2009).
\newblock Can't get there from here: {{The}} decision to apply to a selective
  college.
\newblock {\em Economics of Education Review}, 28(5):620--628.

\bibitem[Groen, 2004]{groen_effect_2004}
Groen, J. A. J.~A. (2004).
\newblock The effect of college location on migration of college-educated
  labor.
\newblock {\em Journal of Econometrics}, 121(1-2):125--142.

\bibitem[Kraft, 2018]{627472}
Kraft, M.~A. (2018).
\newblock Interpreting effect sizes of education interventions.
\newblock {\em Educational Researcher}.

\bibitem[Lehnert et~al., 2020]{lehnert_employment_2020}
Lehnert, P., Pfister, C., and {Backes-Gellner}, U. (2020).
\newblock Employment of {{R}}\&{{D}} personnel after an educational supply
  shock: {{Effects}} of the introduction of {{Universities}} of {{Applied
  Sciences}} in {{Switzerland}}.
\newblock {\em Labour Economics}, 66:101883.

\bibitem[Long, 2008]{long_college_2008}
Long, M.~C. (2008).
\newblock College quality and early adult outcomes.
\newblock {\em Economics of Education Review}, 27(5):588--602.

\bibitem[Marta~Ferreyra et~al., 2017]{marta_ferreyra_at_2017}
Marta~Ferreyra, M., Avitabile, C., Botero~Alvarez, J., Haimovich~Paz, F., and
  Urz{\'u}a, S. (2017).
\newblock {\em At a {{Crossroads}}: {{Higher Education}} in {{Latin America}}
  and the {{Caribbean}}}.
\newblock Directions in {{Development}} - {{Human Development}}. {The World
  Bank}.

\bibitem[{Minist{\'e}rio da Educa{\c c}{\~a}o},
  2012]{ministerio_da_educacao_relatorio_2012}
{Minist{\'e}rio da Educa{\c c}{\~a}o} (2012).
\newblock {Relat\'orio da Comiss\~ao Constitu\'ida pela Portaria
  n\textordmasculine{} 126/2012 sobre a An\'alise da Expans\~ao das
  Universidades Federais \textendash{} 2003 a 2012 \textendash{} Andifes}.

\bibitem[Pereira et~al., 2019]{pereira_geobr_2019}
Pereira, R., Gon{\c c}alves, C., {\noopsort{araujo}}de Araujo, P., Carvalho,
  G., Nascimento, I., and {\noopsort{arruda}}de Arruda, R. (2019).
\newblock Geobr: An {{R}} package to easily access shapefiles of the
  {{Brazilian Institute}} of {{Geography}} and {{Statistics}}.
\newblock Technical report, {IPEA}.

\bibitem[Silva et~al., 2018]{silva_valor_2018}
Silva, C. A.~T., {\noopsort{brito}}de Brito, A. d.~M., {\noopsort{brito}}de
  Brito, A. d.~M., and Faria, J. L.~F. (2018).
\newblock {Valor pago por aluno adicional nas universidades federais
  brasileiras com o programa REUNI}.
\newblock {\em Revista da CGU}, 10(16):22.

\bibitem[Spence, 1973]{spence_job_1973}
Spence, M. (1973).
\newblock Job {{Market Signaling}}.
\newblock {\em The Quarterly Journal of Economics}, 87(3):355.

\bibitem[Spiess and Wrohlich, 2010]{spiess_does_2010}
Spiess, C.~K. and Wrohlich, K. (2010).
\newblock Does distance determine who attends a university in {{Germany}}?
\newblock {\em Economics of Education Review}, 29(3):470--479.

\bibitem[Toivanen and V{\"a}{\"a}n{\"a}nen, 2016]{toivanen_education_2016}
Toivanen, O. and V{\"a}{\"a}n{\"a}nen, L. (2016).
\newblock Education and {{Invention}}.
\newblock {\em The Review of Economics and Statistics}, 98(2):382--396.

\bibitem[Vilela et~al., 2017]{vilela_as_2017}
Vilela, L., Tachibana, T.~Y., Filho, N.~M., and Komatsu, B. (2017).
\newblock {As cotas nas universidades p\'ublicas diminuem a qualidade dos
  ingressantes?}
\newblock {\em Estudos em Avalia\c{c}\~ao Educacional}, 28(69):652--684.

\bibitem[{World Bank}, 2018]{world_bank_world_2018}
{World Bank} (2018).
\newblock World {{Bank Education Overview}}: {{Higher Education}}.
\newblock Technical report.

\bibitem[{World Bank}, 2020]{world_bank_school_2020}
{World Bank} (2020).
\newblock School enrollment, tertiary (\% gross) | {{Data}}.
\newblock
  https://data.worldbank.org/indicator/SE.TER.ENRR?end=2020\&start=1970\&view=chart.

\end{thebibliography}

% ===========================================================

\end{document}